\documentclass[a4paper,12pt]{article}
\usepackage{soul}
\usepackage{ulem}
\usepackage[usenames,dvipsnames]{color}
\usepackage{jheppub}
\usepackage{amsmath, amssymb, slashed, epsf, color, graphicx, latexsym}
\usepackage{epsfig}
\usepackage{graphics}

\newcommand{\hnabla}{\hat{\nabla}}

\begin{document}
	
	\title{General Theory of Large $D$ Membranes Consistent with Second Law of Thermodynamics}
	\author{Arunabha Saha}
	\affiliation{University of Geneva, 24 quai Ernest-Ansermet, 1211 Geneve 4, Switzerland}
	\emailAdd{arunabha.saha@unige.ch}

	\abstract{ We write down the most general membrane equations dual to black holes for a general class of gravity theories, up to sub-leading order in $1/D$ in large $D$ limit.~We derive a ``minimal" entropy current which satisfies a local form of second law from these membrane equations.~We find that consistency with second law requires the membrane equations to satisfy certain constraints. We find additional constraints on the membrane equations from the existence of membrane solutions dual to stationary black holes.~Finally we observe a tension between second law and matching with Wald entropy for dual stationary black hole configurations, for the minimal entropy current. We propose a simple modification of the membrane entropy current so that it satisfies second law and also the stationary membrane entropy matches the Wald entropy.}

\maketitle
\section{Introduction}
Black hole dynamics simplifies considerably in the limit in which dimension of space-time ($D$) tends to infinity while retaining a large isometry. In this limit the dynamics of black holes is dual to the dynamics of a much simpler non-gravitational system.~The dynamics of this dual non-gravitational system is given by a set of effective equations on a finite number of variables in a $1/D$ expansion. In a formalism developed in \cite{Bhattacharyya:2015dva, Bhattacharyya:2015fdk, Dandekar:2016fvw, Bhattacharyya:2017hpj, Bhattacharyya:2018szu, Saha:2018elg, Kundu:2018dvx, Kar:2019kyz, Dandekar:2019hyc}, the effective non-gravitational system consists of a co-dimension one membrane moving in an ambient space-time that is equivalent to the asymptotic space-time of the dual black holes\footnote{There exists another equivalent formalism where the non-gravitational system is defined by a mass and momentum density \cite{Emparan:2015hwa, Emparan:2015gva}. Both these formalism give results consistent with each other in many scenarios where their results have been compared \cite{Dandekar:2016jrp, Mandlik:2018wnw}. See \cite{Emparan:2020inr} for a review of recent progresses mostly based on this formalism}. In this paper we will make some general observations about the effective equations of these dual membranes for a general class of gravity theories. 

In \cite{Bhattacharyya:2016nhn} it was shown that the large $D$ membrane equations derived in \cite{Dandekar:2016fvw} capture the second law of thermodynamics of the dual black hole for Einstein-Hilbert gravity, without it being used as an input.  From this one can conclude that the membrane equation for a more general theory of gravity can be used to obtain a candidate for black hole entropy for this theory which satisfies the second law of thermodynamics. 

It is well known that though Wald entropy correctly captures the first law of black hole thermodynamics, it fails to satisfy the second law. Appropriate corrections to Wald entropy which satisfy the second law can be written down for general theories of gravity in situations with a restricted class of dynamics. e.g. in \cite{Bhattacharjee:2015yaa, Wall:2015raa} a correction to Wald entropy is obtained for small amplitude dynamics around stationary black hole configurations up to linear order in the deviation. The authors of this paper used the physical process version of first law to arrive at this correction. The authors of \cite{Bhattacharya:2019qal} then showed that the requirement of physical process first law can be relaxed by correctly capturing the presence of a spacial entropy current on the horizon along with the entropy density derived in \cite{Bhattacharjee:2015yaa, Wall:2015raa}. In a different context the authors of \cite{Bhattacharyya:2016xfs} found a correction to the Wald entropy which satisfies the second law for non-linear but spherical dynamics of black holes. But there is still no known candidate of entropy which satisfies both first and second law for general dynamics of black holes in general gravity theories. 

In \cite{Dandekar:2019hyc} we initiated a study wherein we looked for a candidate entropy which satisfies second law using the large $D$ limit for Einstein-Gauss-Bonnet theory of gravity. We found a membrane entropy current which satisfies the second law up to linear order in the Gauss-Bonnet (GB) parameter for general non-linear dynamics in the large $D$ limit. This entropy current also matched the Wald entropy up to leading order in large $D$ though it was not used as an input for the derivation. These results indicate that the large $D$ membrane/black-hole duality is an interesting set up to gain a better understanding of the second law of black hole thermodynamics for  theories of gravity beyond Einstein-Hilbert gravity and for configurations with more general non-linear dynamics. 

To work towards this goal, in this paper we write down the membrane equations dual to large $D$ black hole dynamics for general higher derivative theories of gravity, upto first sub-leading order in $1/D$. Using this general membrane equations we then present a systematic algorithm to obtain a membrane entropy current for these general gravity theories, demanding that this entropy current satisfies a local form of second law. Our formalism applies to those theories of gravity which have a smooth limit to Einstein gravity. We demonstrate that the membrane equations need to satisfy certain constraints for it to be consistent with the local form of second law. In addition we further constrain the form of the membrane equations by demanding the presence of membrane configurations dual to stationary black holes. 

In our construction we do not a priori demand that the membrane entropy for the stationary configurations match with the Wald entropy for corresponding black holes. We find that for the minimal version of entropy current derived by us, matching with Wald entropy can be used as an additional consistency condition for the membrane equations and the construction of entropy current. But for a non-perturbative analysis  matching with Wald entropy at leading order in $1/D$ is in tension with the second law. We then show that this tension can be easily ameliorated by a simple modification of the minimal entropy current.

In the next subsection we start our analysis by explaining in details the general higher derivative theories of gravity that we will work with.

\section{The gravity action and various length scales}
 We work with classical gravity theories for which the equations of motion are defined in  terms of the metric of the space-time and its derivatives. We only consider theories with vanishing matter stress tensor and cosmological constant. In addition we demand that the theories under consideration have a smooth limit to Einstein-Hilbert theory of two derivative gravity. Also, we only consider those black hole solutions which are continuously connected to  solutions of Einstein-Hilbert gravity in the limit in which the couplings determining the higher derivative terms tend to zero.  

We group together terms in the gravity equation with the same total number of derivatives acting on the metric. Though the number of metric fields in these terms need not be same.  The schematic form of the equation of these gravity theories is 
 \begin{equation}
 	E^{(EH)}_{MN}+\sum_{n=3}\tilde{\alpha}^{2m}E^{n}_{MN}=0
 \end{equation}
where, $E^{(n)}_{MN}$ denotes terms in the gravity equation with $n$ number of derivatives acting on the metric.~$\tilde{\alpha}$ is the length scale below which the higher derivative terms in the gravity equation are of the same order as the Einstein-Hilbert part of the equation. In other words for metrics with derivatives of the order of $\mathcal{O}(1/\tilde{\alpha})$, the gravity equation has comparable contributions from the Einstein-Hilbert part and the higher derivative parts. 

 The gravity equation is a tensor with two free indices. Hence, if there are `x' derivatives and `y' metric fields in a given term of the gravity equation then $x+2y-2$ of the indices have to be contracted among themselves. Therefore, $x+2y$ is an even positive integer and hence $x$ is also an positive even number. So, each term in the gravity equations can only have an even number of derivatives acting on the metric fields, i.e. $n=2m+2$ with $m\ge1$. 

We will confine our attention to black hole dynamics in the large $D$ limit where, the derivative of the metric away from the horizon is $\mathcal{D}$ and the transverse derivatives (including those along time like directions )are $\mathcal{O}(D^)$. This is expected to be the characteristics of the late time dynamics of black holes at times scales of the order of $\mathcal{O}(1/L_H)$ where $L_H$ is the horizon length scale which we assume to be of $\mathcal{O}(D^0)$. This expectation is borne out by the quasi-normal mode analysis at large $D$ for Einstein-Hilbert gravity \cite{Emparan:2014aba}. This also holds true for static black holes in general theories of gravity. e.g. the static black hole in Schwarzschild coordinates for Einstein-Hilbert gravity has the metric 
$$-dt^2\left(1-\left(\frac{r_h}{r}\right)^{D-3}\right)+\frac{dr^2}{\left(1-\left(\frac{r_h}{r}\right)^{D-3}\right)}+r^2d\Omega_{D-2}^2$$
 where, the coefficients of both the $dt^2$ and $dr^2$ term contain the blackening factor which has $\mathcal{O}(D)$ derivatives along the $r$ direction. Therefore, if the length scale $\tilde{\alpha}$ is such that $\frac{\tilde{\alpha}}{L_H}\simeq\mathcal{O}(D^0)$, in the large $D$ limit, the higher derivative terms will dominate over the Einstein-Hilbert terms in the gravity equations for black hole solutions.~Since, we consider only those solutions of the gravity equations which are continuously connected to the solutions of  Einstein-Hilbert gravity, we impose a `kinematic constraint' on the length scale $\alpha$ given by
\begin{equation}
\frac{\tilde{\alpha}}{L_H}=\mathcal{O}(1/D) 
\end{equation}
so that the higher derivative terms contribute at the same or lower order in $1/D$ compared to the Einstein-Hilbert term in the gravity equations. 

For static black hole under the black hole/membrane duality in the large $D$ limit the radius of the horizon $r_H$ in the black hole side is related to the trace of the extrinsic curvature of the dual membrane ($\mathcal{K}$) as $$\mathcal{K}=\frac{D}{r_H}.$$
For dynamical black holes where the dynamics is confined to finite number of directions ( which is the situation of interest to us), the map between the membrane curvature and the local horizon length scale is of the form $\mathcal{K}=\frac{D}{l_H}$. Using this we  can define a dimensionless constant $\alpha_K$ given by 
\begin{equation}
	\boxed{\alpha_K=\tilde{\alpha} \mathcal{K}}
\end{equation}
which makes sure that $\frac{\tilde{\alpha}}{L_H}\simeq\mathcal{O}(D^{-1})$ since $\mathcal{K}\simeq \frac{D}{L_H}$. $\alpha_K$ is the parameter which when tuned to zero, the theory goes over to the Einstein Hilbert gravity and the solutions tend to those in Einstein-Hilbert theory. 
\section{ Towards a strategy to find the most general entropy current}
We now sketch out the broad strategy that we follow to derive a membrane entropy current for a general class of membrane equations. To arrive at the strategy it is worthwhile to review the derivation of the entropy current for EGB gravity presented in \cite{Dandekar:2019hyc} in manner which sheds light on the strategy one needs to follow in a more general situation. 

\subsection{Construction of entropy current for EGB gravity up to linear order in $\beta$}
In the notation used in \cite{Dandekar:2019hyc} the Gauss Bonnet (GB) parameter $\beta$ is related to $\alpha_K$ by 
\begin{equation}
	\frac{\beta\mathcal{K}^2}{D^2}=\alpha_K^2
\end{equation}
The scalar membrane equation for Einstein-Gauss-Bonnet (EGB) gravity up to linear order in the GB parameter ($\beta$) was obtained in \cite{Dandekar:2019hyc}, and is given by
\begin{equation}\label{sca_eqn}
	\hnabla\cdot u=\frac{2}{\mathcal{K}}\left(1-\frac{\beta\mathcal{K}^2}{D^2}\right)\sigma_{\alpha\beta}\sigma^{\alpha\beta}+\frac{4\beta\mathcal{K}}{D^2}\sigma_{\alpha\beta}K^{\alpha\beta}+\frac{8\beta}{\mathcal{K}D^2}\hnabla^\alpha\sigma_{\alpha\beta}\hnabla^\gamma\sigma_\gamma^\beta
\end{equation}
Here $\hnabla$ denotes covariant derivatives w.r.t the induced metric in the membrane world-volume. In the limit of $\alpha_K\rightarrow 0$ the membrane equation matches that of Einstein-Hilbert gravity. Hence, we start with the following form of entropy current for this scalar membrane equation, 
\begin{equation}
	J^\mu_S=\frac{u^\mu}{4}+\beta \mathcal{J}^\mu
\end{equation}
which approaches the entropy current for Einstein-Hilbert gravity derived in \cite{Bhattacharyya:2016nhn} in the $\alpha_K\rightarrow 0$ limit. . One can arrive at an entropy current $J_S$ by choosing suitable terms to add to $\mathcal{J}^\mu $ which cancel out the non-positive definite terms in $\nabla\cdot u/4$. This procedure, if successful, by construction will produce an entropy current with positive definite divergence.

The first term of \eqref{sca_eqn} is positive definite since $|\beta|\ll1$. The second term in \eqref{sca_eqn} is definitely not positive definite and the third term is negative definite for $\beta< 0$ (Dynamics with large isometry in the large $D$ limit have $\mathcal{K}\ge0$). The following expression of $\mathcal{J}^\mu$ takes care of the second term in \eqref{sca_eqn}.
\begin{equation}
	\mathcal{J}^\mu=-\frac{\mathcal{K}}{D^2}(u\cdot K)^\mu +\mathcal{F}^\mu
\end{equation}
Taking the divergence of the above expression we get
\begin{eqnarray}
	\hnabla\cdot \mathcal{J}&=&-\frac{\mathcal{K}}{D^2}u\cdot\hnabla\mathcal{K}-\frac{\mathcal{K}}{D^2}\hnabla_\mu u_\alpha K^{\alpha\mu}-\frac{\hnabla_\mu\mathcal{K}}{D^2}(u\cdot K)^\mu+\hnabla\cdot \mathcal{F}
\end{eqnarray}
Substituting the following decomposition of action of covariant derivative on velocity vector in the above expression, 
\begin{eqnarray}
	\hnabla_\mu u_\alpha&=&\sigma_{\mu\alpha}+\omega_{\mu\alpha}+a_\alpha u_\mu +\frac{\mathcal{P}_{\mu\alpha}}{D-2}\hnabla\cdot u\nonumber\\
	\text{where,}\quad \sigma_{\mu\alpha}&=&\mathcal{P}^\beta_\mu \mathcal{P}^\gamma_\alpha\left(\frac{\hnabla_\beta u_\gamma+\hnabla_\gamma u_\beta}{2}\right),\quad 	\omega_{\mu\alpha}=\mathcal{P}^\beta_\mu \mathcal{P}^\gamma_\alpha\left(\frac{\hnabla_\beta u_\gamma-\hnabla_\gamma u_\beta}{2}\right),\quad a_\gamma=(u\cdot\hnabla u)_\gamma \nonumber\\
	\text{and,}&&\mathcal{P}_{\mu\alpha}=g_{\mu\alpha}+u_\mu u_\alpha
\end{eqnarray}
and observing that the membrane equation imply that $\hnabla\cdot u=\mathcal{O}(1/D)$, we get
\begin{equation}
	\hnabla\cdot \mathcal{J}=-\frac{\mathcal{K}}{D^2}u\cdot \hnabla\mathcal{K}-\frac{\mathcal{K}}{D^2}\sigma_{\mu\alpha}K^{\alpha\mu}-\frac{\mathcal{K}}{D^2}(u\cdot \hnabla u)_\alpha(u\cdot K)^\alpha-\frac{\hnabla_\mu\mathcal{K}}{D^2}(u\cdot K)^\mu+\hnabla\cdot \mathcal{F}
\end{equation}
We see that though the $\mathcal{J}^\mu$ written above removes one of the non-positive definite terms from the divergence of the entropy current, but it comes at the cost of addition of the first term in the above expression which is at a higher order in $D$ as compared to the Einstein-Hilbert entropy production. Also, this term is not positive definite. Once again this can corrected by adding the following $\mathcal{O}(D^0)$ term in the expression of $\mathcal{F}^\mu$
\begin{eqnarray}
	\mathcal{F}^\mu=\frac{\mathcal{K}^2}{2D^2}u^\mu+\mathcal{S}^\mu.
\end{eqnarray}
Again using the fact that $\hnabla\cdot u$ is $\mathcal{O}(1/D)$ along with the above expression for $\mathcal{F}^\mu$ we get
\begin{equation}
	\hnabla\cdot \mathcal{J}=-\frac{\mathcal{K}}{D^2}\sigma_{\mu\alpha}K^{\alpha\mu}-\frac{\mathcal{K}}{D^2}(u\cdot \hnabla u)_\alpha(u\cdot K)^\alpha-\frac{\hnabla_\mu\mathcal{K}}{D^2}(u\cdot K)^\mu+\frac{\mathcal{K}^2}{D^2}\hnabla\cdot u+\hnabla\cdot \mathcal{S}
\end{equation}
Using the above expression in the divergence of the entropy current along with the scalar membrane equation (and keeping terms only to linear order in $\beta$) we get
\begin{eqnarray}
	\hnabla\cdot J_S&=&\frac{1}{2\mathcal{K}}\left(1+\frac{\beta\mathcal{K}^2}{D^2}\right)\sigma_{\alpha\beta}\sigma^{\alpha\beta}+\frac{2\beta}{\mathcal{K}D^2}\hnabla^\alpha\sigma_{\alpha\beta}\hnabla^\gamma\sigma^\beta_\gamma-\beta \frac{\hnabla_\mu\mathcal{K}}{D^2}(u\cdot K)^\mu\nonumber\\
	&&-\frac{\beta\mathcal{K}}{D^2}u\cdot(\hnabla u)_\alpha (u\cdot K)^\alpha +\beta\hnabla\cdot \mathcal{S}\nonumber\\
	\text{where,} \quad J_S^\mu&=&\frac{u^\mu}{4}+\beta \left(-\frac{\mathcal{K}}{D^2}(u\cdot K)^\mu+\frac{\mathcal{K}^2}{2D^2}u^\mu+\mathcal{S}^\mu\right)
\end{eqnarray}
We see that the RHS is still not positive definite for arbitrary sign of $\beta$. The appropriate expression for $\mathcal{S}$ which will take care of all of the problematic terms in the RHS is
\begin{equation}
	\mathcal{S}^\mu=-\frac{2}{\mathcal{K}D^2}\hnabla^\alpha\sigma_{\alpha\beta}\sigma_\mu^\beta+\frac{1}{D^2}u\cdot(\hnabla u)^\alpha\hnabla_{\alpha}u^\mu+\frac{1}{\mathcal{K}D^2}\hnabla^\alpha\mathcal{K}\hnabla_{\alpha}u^\mu
\end{equation}
To see that the above expression of $\mathcal{S}$ gets rid of the non positive definite terms in divergence of entropy current we compute the divergence of $\hnabla_\alpha u^\mu$ as
\begin{eqnarray}
	\hnabla^\mu\left(\hnabla_{\alpha} u_\mu\right)&=&R^\mu_{\theta\mu\alpha}u^\theta +\hnabla_{\alpha}(\hnabla\cdot u) \nonumber\\
	&=& \mathcal{K}(u\cdot K)_\alpha +\mathcal{O}(1/D).
\end{eqnarray}
In the first line above $R^\mu_{\theta\mu\alpha}$ is the Riemann tensor of the induced metric on the membrane. In the second line we use Gauss-Codacci relations and finally we arrive at the following expression of membrane entropy current which has a positive definite divergence for all allowed membrane configurations. 
\begin{eqnarray}
	&&J_S^{\mu}=\frac{u^\mu}{4}+\beta \Bigg(-\frac{\mathcal{K}}{D^2}(u\cdot K)^\mu+\frac{\mathcal{K}^2}{2D^2}u^\mu-\frac{2}{\mathcal{K}D^2}\hnabla^\alpha\sigma_{\alpha\beta}\sigma_\mu^\beta+\frac{1}{D^2}u\cdot(\hnabla u)^\alpha\hnabla_{\alpha}u^\mu+\frac{1}{\mathcal{K}D^2}\hnabla^\alpha\mathcal{K}\hnabla_{\alpha}u^\mu\Bigg)\nonumber\\
	&&\text{and,}\quad \hnabla\cdot J^S=\frac{1}{2\mathcal{K}}\left(1+\frac{\beta\mathcal{K}^2}{D^2}\right)\sigma_{\alpha\beta}\sigma^{\alpha\beta}
\end{eqnarray}
From the above exercise we see that the process of finding an entropy current involved adding the smallest number of ``counter-terms" to the expression of the entropy current so that the using the scalar membrane equation, the divergence of this entropy current becomes positive definite. This entropy current by no means is unique, since we can always add terms to it whose divergence is positive definite (still preserving the Einstein-Hilbert limit). 
\subsection{Basic strategy to find the most general entropy current}
We now lay out the basic strategy to write down an entropy current given a general membrane equation.~The entropy current that we write down should reduce to the membrane entropy current for Einstein-Hilbert gravity in the limit of $\alpha_K\rightarrow 0$. i.e the general form of the entropy current is given by
\begin{eqnarray}
	&&J^\mu_S=\frac{u^\mu}{4}+\mathcal{F}^\mu(\alpha_K)\nonumber\\
	\text{where,}\quad &&\lim_{\alpha_K\rightarrow 0} \mathcal{F}^\mu(\alpha_K)=0
\end{eqnarray}
Like in the analysis of EGB gravity, the role of $\mathcal{F}^\mu$ is to take care of terms which are not positive definite in the expression of $\hnabla\cdot J_S$ once the membrane equations have been used up. We will call the terms that we add to $\mathcal{F}^\mu$ to get a consistent entropy current  as the ``counter-terms". 
One may argue that the coefficient of the $\frac{u^\mu}{4}$ term can itself get modified to $\xi(\alpha_K)$ where,  $\lim_{\alpha_K\rightarrow 0}\xi(\alpha_K)\rightarrow 1$, but we can always write $$\xi(\alpha_K)=1+(\xi(\alpha_K)-1)$$ and absorb the $(\xi(\alpha_K)-1)\frac{u^\mu}{4}$ part in the definition of $\mathcal{F}^\mu(\alpha_K)$. In the derivation of entropy current for EGB gravity we had to add a term $\frac{\beta\mathcal{K}^2}{2D^2}u^\mu$ which is an example of a term proportional to $u^\mu$ in $\mathcal{F}^\mu(\alpha_K)$ with $\xi(\alpha_k)=1+\frac{2\beta \mathcal{K}^2}{D^2}$. 
 
 With this convention the algorithm that we will follow to arrive at the minimal membrane entropy current dual to black holes in a general theory of gravity, involves the following steps
\begin{itemize}
	\item First we write down the most general scalar membrane equation at a given order in $1/D$. These equations are expressed in terms of independent variables obtained by solving for the membrane equations at the preceding order in $1/D$. The membrane equations should reduce to the membrane equations for Einstein-Hilbert gravity in the limit of $\alpha_K\rightarrow 0$.  
	\item Next for the terms in the expression of $\hnabla\cdot u$ in the membrane equation which are not positive definite, we add counter-terms to  $\mathcal{F}^\mu(\alpha_K)$ to make $\hnabla\cdot J_S$ positive definite.   
	\item For terms in the expression of $\hnabla\cdot u$ which are not positive definite and for which it is also not possible to write a counter-term, we will need to put adequate constraints on their coefficient in the membrane equations, so that the entropy current is consistent. These coefficients in the membrane equation are functions of $\alpha_K$, and borrowing notation from fluid dynamics, we call these coefficients the ``transport coefficients".  
	\item We also impose constraints on the transport coefficients from the existence of membrane configurations dual to stationary black holes. These constraints in turn put constraints on the expression of the entropy current itself. 
	\item And lastly we demand that for stationary solutions the membrane entropy obtained by integrating the entropy current on space-like slices in the membrane world-volume matches with the Wald entropy of the corresponding dual black hole. 
\end{itemize}

In the next section we elaborate on the procedure to find out all possible scalar objects made out of membrane shape and velocity field, that can appear in the scalar membrane equation up to the sub-leading order in $1/D$. 
\subsection{General structure of the scalar membrane equation}
The schematic form of the scalar membrane equation is given by
$$\nabla\cdot u=\mathcal{M}$$
where, $\mathcal{M}$ is an $\mathcal{O}(D^0)$ or higher in $1/D$ scalar function composed of various terms which are functions of the shape and velocity functions of the membrane. The length dimension of the left hand side  is $[L]^{-1}$ and hence the RHS should also have the same dimensions. 

As an illustration if types of objects that can contribute to the RHS above, let us consider the contribution to $\mathcal{M}$ of two scalars formed out of the shape and velocity functions  : $1)$ $(u\cdot K\cdot u)$ and $2)$ $\sigma_{\mu\nu}\sigma^{\mu\nu}$. The dimension of $u\cdot K\cdot u$ is $[L]^{-1}$ and the dimension of  $\sigma_{\mu\nu}\sigma^{\mu\nu}$ is $[L]^{-2}$. Hence, we need to multiply the second object with another object with the dimension of $[L]$ before adding it to $\mathcal{M}$.  
Whenever, we need to adjust the dimension of a quantity to be added into the scalar membrane equation we will multiply it with the adequate powers $\mathcal{K}$, because it provides a natural object with a length scale associated with it in the membrane. 

After that depending on the order in $1/D$ at which these objects are contributing to the scalar membrane equation we multiply the objects with additional powers of $D$. e.g if both the quantities mentioned here contribute to the scalar membrane equation at $\mathcal{O}(1/D)$, then they are added to the RHS in the following manner. 
$$\nabla\cdot u=A(\alpha_K)\frac{u\cdot K\cdot u}{D}+B(\alpha_K)\frac{\sigma_{\mu\nu}\sigma^{\mu\nu}}{\mathcal{K}}$$
where, $A(\alpha_K)$ and $B(\alpha_K)$ are the ``transport coefficients" associated with these two terms. These transport coefficients are functions of the dimensionless quantity $\alpha_K$ with the appropriate $\alpha_K\rightarrow 0$ limit.

Hence to find all possible terms in the expression od $\mathcal{M}$ we need to list together all possible scalar objects that can contribute at relevant order and then multiply them with appropriate powers of $\mathcal{K}$ and $D$ to adjust length dimensions and orders in $1/D$. 

To arrive at the exhaustive list of objects that can contribute to the membrane equations we need to analyse the mechanism by which the gravity equations give rise to the membrane equations in general. Next we present a brief outline of this procedure drawing heavily from the derivation of the membrane equations for different theories with two or more derivatives in \cite{Bhattacharyya:2015dva, Bhattacharyya:2015fdk, Dandekar:2016fvw, Bhattacharyya:2017hpj, Bhattacharyya:2018szu, Saha:2018elg, Kar:2019kyz, Dandekar:2019hyc}.
\subsection{An outline of the derivation of large $D$ membrane equations for the general gravity equations}
 In the large $D$ regime we solve for the metric of dynamical black hole perturbatively, order by order in $1/D$. The starting point for this procedure is to assume that the theory of gravity under consideration has a static black hole solution. Then the static black hole metric in Kerr-Schild coordinates is used to construct the ansatz metric to start the perturbative procedure. For a general theory of gravity the static asymptotically flat black hole metric in the Kerr-Schild coordinates has the form 
\begin{eqnarray}
	ds^2&=&-dt^2+dr^2+r^2d\Omega_{D-2}^2+F\left(r\right)(dt+dr)^2\nonumber\\
		\text{where,} \quad &&F(r=r_h)=1 \quad \lim_{r\rightarrow\infty}F(r)=0\nonumber
\end{eqnarray}
where $r=r_h$ is the position of the horizon In these coordinates. In the  large $D$ limit the blackening factor $F(r)$ satisfies the following additional properties 
$$ \partial_r F=\mathcal{O}(D) \quad \text{for,}\quad \frac{r}{r_h}-1=\mathcal{O}(1/D)$$
and, $$F(r)\simeq 0\quad \text{for,}\quad \frac{r}{r_h}-1=\mathcal{O}(D^0).$$
For static black holes of Einstein-Hilbert gravity, $F(r)=\left(\frac{r_h}{r}\right)^{D-3}$. For examples of blackening factor in other situations see \cite{Bhattacharyya:2015fdk}(for Einstein-Maxwell theory), \cite{Saha:2018elg} (for Einstein-Gauss Bonnet theory) and \cite{Kar:2019kyz}( for general four derivative theory of gravity). In general $F(r)$ is obtained by solving an ordinary differential equation (ODE) along $r$ for $F(r)$ with the above boundary conditions . This ODE depends on the theory of gravity under consideration. The above metric can be written in a covariant form as,
\begin{equation}
	ds^2=ds_{flat}^2+F(\psi)(O_M dx^M)^2
\end{equation}
where,$$\psi=\frac{r}{r_h}=1\quad\text{and,}\quad O_M dx^M=n_M dx^M-u_M dx^m\quad \text{where,}\quad n_M dx^M=dr, \quad u_M dx^M=-dt$$
It is easy to see that the expressions for $\psi$ and $u_M dx^M$ written above satisfy the following equations,
$$ \nabla^2\left(\frac{1}{\psi^{D-3}}\right)=0\quad \text{and,}\quad \nabla\cdot u=0$$
The covariant derivatives and dot products are w.r.t $ds_{flat}^2$. The above metric can be thought of as being dual to a membrane embedded in flat space-time where, $\psi=1$ is its position and $u^M\partial_M$ (which is also the generator of the event horizon) can be thought of as a  velocity field in its world-volume (as it is orthogonal to the normal to the membrane). 

If the shape ($\psi$) and velocity vector ($u_M$) are elevated to arbitrary functions of the space-time coordinates with $\mathcal{O}(D^0)$ derivatives with the following properties
\begin{eqnarray}\label{membrane_conditions}
	&&O_M=n_M-u_M,\quad n_M=\frac{d\psi}{\sqrt{d\psi\cdot d\psi}}, \quad u\cdot u=-1,\quad u\cdot n=0\nonumber\\
	&&\nabla^2\left(\frac{1}{\psi^{D-3}}\right)=\mathcal{O}(D)\quad \text{and,}\quad \nabla\cdot u=\mathcal{O}(D^0).
\end{eqnarray}
then the covariant form of the static metric in the Kerr-Schild form still satisfies the gravity equations to leading order in large $D$ provided $F(\psi)$ continues to satisfy the the same differential equation and boundary condition as in the static case in the membrane region $\frac{r}{r_h}-1\sim\mathcal{O}(1/D)$ ( see appendix \eqref{Fequation} for details). Hence, this can be taken to be the starting ansatz metric for the perturbative analysis in $1/D$. We still demand that $\psi=1$ continues to be the position of the horizon and $u^\mu$ continues to the the generator of event horizon. The two equations in the last line simply say that both these quantities should evaluate to one order lower in $D$ than their naive order. The second of the two equations can be interpreted as the fact that to leading order $u^M$ is a time like killing vector. All the dot products and covariant derivatives are w.r.t the flat space-time. Also, $F(\psi)$ continues to satisfy the following large $D$ properties. 
 \begin{eqnarray}
 		&& \partial_M F=\partial_M\psi \partial_{\psi}F=\mathcal{O}(D)\quad \text{since,}\quad \partial_\psi F=\mathcal{O}(D)\nonumber\\
 	&&\text{also,}\lim_{\psi\rightarrow\infty}F(\psi)=0\quad \text{and,} 
 \end{eqnarray}

The ansatz metric fails to solve for the gravity equation beyond leading order in $1/D$ for generic functions $\psi$ and $u$ and one needs add corrections to the ansatz metric at sub-leading order in $1/D$ to make them solve the gravity equations at desired order. Because of the presence of the fast direction in the space-time along $d\psi$, it turns out that the metric corrections effectively need to only solve ordinary differential equations along $d\psi$ to arrive at a solution to gravity equation (see \cite{Dandekar:2016fvw, Bhattacharyya:2017hpj} for details). For regions where $\psi-1\gg \mathcal{O}(D)$ the metric decays to that of flat space-time as $e^{-(\psi-1)}$ and by default it solves for the gravity equations \footnote{We assume that all of the gravity equations that we consider have flat space-time as solution}. Hence, our main concern is to solve for the gravity equations in the region where $\psi=1+\frac{R}{D}$ with $R\sim\mathcal{O}(D^0)$.  We call this the ``membrane region" of the metric. The inhomogeneous parts ( source terms) for these differential equations are functions of $R$ and ultra-local quantities formed out of the shape and velocity vector fields about any arbitrary point in the membrane region. 

It turns out that for any arbitrary configuration of the membrane shape and velocity, it is not possible to obtain metric correction that are regular everywhere in the membrane region and at any given order the membrane configuration needs to satisfy certain constraints on the ultra-local quantities in the source terms of the differential equations. These constraints are interpreted as the membrane equations. The presence of these constraints is related to the diffeomorphism degree of freedom of the metric. Due to this the number of independent physical degrees of freedom in the metric are lesser than the number of gravity equations. Some of the gravity equations, more precisely the equations with a component along the $d\psi$ are not dynamical but are rather constraint equations. Hence, once the dynamical equations have been used up to solve for the metric components in a given gauge, the constraint equations give rise to the constraints on the membrane functions. The general gravity equations that we work with are diffeomorphism invariant and hence we expect them to follow the same pattern. 

Hence, if we can list together all possible objects that can contribute to the inhomogeneous parts of the gravity equation we can list all possible terms that can possibly contribute to the membrane equations. In the next section we undertake the task to classify all possible objects in the source terms of the gravity equations. 

 \section{Classification of objects contributing to the inhomogeneous parts of gravity equations}
 The general gravity equation that we consider is diffeomorphism invariant. So, the term $E_{MN}^{(2m+2)}$ with $2m+2$ derivatives in the gravity equation, consists of a string of covariant derivatives $\nabla$ and internal curvature tensors (or scalars) with the following schematic form 
 $$\nabla\ldots \nabla \tilde{R}\ldots\tilde{R}\ldots\nabla\ldots\nabla\tilde{R}\ldots.$$
 where $\tilde{R}$ denotes the internal curvatures of the metric. The indices have been suppressed for convenience and are not important for the argument that we present. Let there be $`A'$ number of $\nabla$s and $`B'$ number of $\tilde{R}$s in the above expression. From the constraint on the total number of derivatives of this term we have
 $$A+2 B=2m+2.$$
 Hence, $A$ is an even number, say $A=2q$. Also since, $q\ge 0$ from the above equation we have 
 $$\boxed{B\le m+1}.$$
 
 To classify all the objects that can contribute to the gravity equations we need to track the number of derivatives and their type, on the $\psi$ and $u$ functions in the metric , in the general gravity equations. We also keep in mind that we will recast all covariant derivatives into derivatives in flat space-time plus the effect of non-trivial structure of the metric. We can classify the different elements of $E_{MN}^{(2m+2)}$ by their contribution to order in $D$. Let
 \begin{enumerate}
 	\item $z$ number of $\tilde{R}$s in the above evaluate to $\mathcal{O}(D^0)$. 
 	\item $w$ of the $\tilde{R}$s evaluate to $\mathcal{O}(D)$
 	\item So, $B-z-w$ of the $\tilde{R}$s evaluate to $\mathcal{O}(D^2)$. 
 	\item x of the $\nabla$s act without increasing any order in $D$. If the covariant derivatives act without contraction with a vector and if they do not act on $F(\psi)$ or its derivative in the $\tilde{R}$s, no factor of $D$ is added.
 	\item $2y$ of the derivatives act as $\nabla^2$ on a scalar or vector in the above string. This action produces a factor of $D$ for each $\nabla^2$.
 	\item So, $2q-x-2y$ of the $\nabla$s produce a factor of $D$ for each $\nabla$. This can happen if $\nabla$ acts on either $F(\psi)$ or its derivatives or acts by contraction on a tensor in the $\tilde{R}$s\footnote{Contraction of covariant derivatives produce extra $D$ due to the large isometry of the dynamics.}. 
 \end{enumerate}
 Hence, taking into account the $D^{-2m}$ in the coefficient of $E_{MN}^{(2m+2)}$ in the gravity equation the net order in $D$ of the above combination is
 $$D^{2-y-x-w-2z}\quad \because B+q=m+1.$$
 with each of $x$, $y$, $z$, $w$ being positive integers and also
 $$\boxed{2q\ge x+2y}\quad \text{and,}\quad \boxed{B\ge z+w}.$$
 The last two inequalities are there because number of $\nabla$s and $\tilde{R}$s in all categories must be integers greater than or equal to zero. 
 The classification of various objects then depends on demanding that the gravity equations contribute at the right order: namely at $\mathcal{O}(D)$ for first sub-leading order equations and $\mathcal{O}(D^0)$ for second sub-leading order equations. We present the details of the result in appendix \eqref{classificationdetails}. We summarise the final result here in terms of objects formed out of membrane vectors $A_M^{(i)}=\{u_M,n_M\}$ and the scalar $\mathcal{Q}=\nabla\cdot n$ where the divergence is w.r.t flat space-time. 
 
 The objects that we consider are obtained by the action of derivatives on $A_M^{(i)}$ and $\mathcal{Q}$. We identify objects which only differ by overall factors of $\mathcal{Q}$ and $A_{M}^{(i)}$ as equivalent. We do so because in the analysis of the independent objects in the membrane equation they will not contribute different information. 
 \subsection{First sub-leading order gravity equation}
 The equation at this order is $x+y+w+2z=1$ and the analysis of this condition in appendix \eqref{classificationdetails} tells us that the objects with maximal number of derivatives acting on $A_M^{(i)}$ and $\mathcal{Q}$ have the following structure
 \begin{equation}
 	\boxed{\nabla^2\mathcal{Q}\quad \text{and}\quad \nabla^2 A_m^{(i)}}
 \end{equation}
i.e. two derivative objects need to appear as $\nabla^2$ and not in any other combination. All the covariant derivatives are w.r.t flat space-time All objects with lower number of derivatives contribute to the gravity equation with arbitrary distribution of free indices. All objects are linear in $A_M^{(i)}$ and $\mathcal{Q}$. 
\subsection{Second sub-leading order gravity equation}
Objects linear in $A_M^{(i)}$ and $\mathcal{Q}$ have maximum of $4$ derivatives acting on them with the following structure
\begin{eqnarray}
	\boxed{\nabla^2\nabla^2\mathcal{Q}\quad \text{and}\quad \nabla^2\nabla^2 A_M^{(i)}}
\end{eqnarray}
The bilinear objects also have a maximum of four derivatives with the structure
\begin{equation}
	\boxed{\nabla^2\mathcal{Q}\nabla^2 \mathcal{Q},\quad \nabla^2\mathcal{Q}\nabla^2 A_M^{(i)}\quad \text{and} \quad \nabla^2 A_m^{(i)}\nabla^2 A_N^{(j)}}
\end{equation}
Both the linear and bi-linear objects with three derivatives are also constrained to have the following structures only
\begin{equation}
	\nabla_M\nabla^2 A_N^{(i)}\quad \text{and,}\quad \nabla_m\nabla^2\mathcal{Q}\nonumber
\end{equation}
\begin{equation}
	\nabla_M\mathcal{Q}\nabla^2\mathcal{Q}, \quad\nabla_M A_N^{(i)}\nabla^2 A_P^{(j)}, \ldots
\end{equation}
i.e. two of the three derivatives have to act as $\nabla^2$. All possible lower derivative objects with arbitrary distribution of free indices is allowed without any further constraints. 

Having classified the objects that can contribute to the gravity equation at different orders in large $D$, we now explain the mechanism to arrive at the membrane equations from the gravity equations. 
\section{From constraint gravity equations to membrane equations}
As we have mentioned in an earlier section, the membrane equation are obtained by evaluating the constraint gravity equation after the dynamical gravity equations are solved. More precisely the membrane equations can be obtained by evaluating the constraint equations along a distinct $\psi=constant$ slice \footnote{By property of the constraint equations the information contained by them is independent of the choice of slice}. We can impose boundary condition on the metric correction so that $\psi=1$ remains the horizon at all orders. This slice is found to be a convenient choice for the task (see\cite{Dandekar:2016fvw, Saha:2018elg} etc). And hence we evaluate the membrane equation on this slice.

There is one final step where we need to process the equation obtained by evaluating the constraint gravity equation at $\psi=1$ a bit more.  The objects  in the gravity equation contain derivatives of the membrane quantities ($A_M^{(i)},\mathcal{Q}$) along all space-time directions including the direction away from $\psi=1$ surface. But since, the membrane data ( including its velocity) is contained solely along the $\psi=1$ surface, we have to find a way to express the normal derivatives in terms of quantities which can be completely defined on the $\psi=1$ surface.  

This can be done by defining a family of surface of which $\psi=1$ surface is a member. A particular convenient choice for this was defined in \cite{Bhattacharyya:2015dva, Dandekar:2016fvw}. There a scalar field $B$ was introduced which is the proper distance from the membrane along a space-like geodesic `shot' from any point on the membrane surface\footnote{We need not worry about caustics of the family of geodesic `shot' from nearby point with distance of the order of unity since, the caustics can possibly appear at distance of the order of average inverse curvature at that point in the membrane. Average inverse curvature is of the order of $\frac{\mathcal{K}}{D}$ which for the us is $\mathcal{O}(D^0)$ and hence is outside the membrane region, where anyway the solution is equivalent to flat space-time.}. This definition makes it clear that the surface of concern is located at $B=0$. Using this definition it can be shown (see\cite{Bhattacharyya:2015dva}) that the normalised normal vector can be written as 
\begin{equation}
	n_M=\partial_MB
\end{equation}
And in \cite{Dandekar:2016fvw} it was shown that the $\psi$ function which satisfies the harmonic equation on $\psi^{-(D-3)}$ (required by the solubility of the leading order ansatz metric) is given in terms of $B$ by
\begin{equation}
	\psi=1+\frac{\mathcal{Q} B}{D}
\end{equation}
where $\mathcal{Q}=\nabla_M n^M$. The above choice of the family of surfaces leads to the following condition on the normal derivative of the normal vector ($n_M=\frac{\partial_M\psi}{\sqrt{d\psi\cdot d \psi}})$
\begin{equation}\label{normderivnormal}
	n\cdot\nabla n_M=0
\end{equation}
Similarly, we can define the velocity field on any member of these family of surfaces by the condition that it continues to be a unit normalised time like vector orthogonal to normal vector and that it is parallel transported along the normal vector. 
$$n\cdot\nabla u_M=0$$
So, we have achieved our target of re-expressing the normal derivatives in terms of quantities defined only on the $\psi=1$ surface by  in a simple manner by equating them to zero. 
For this definition of $\psi$  the extrinsic curvature at any point in the `membrane region' ( $\psi-1\simeq \mathcal{O}(D^{-1})$), defined by the lie derivative of the induced metric of the surface along the normal direction is given by \cite{Dandekar:2016fvw}
\begin{equation}
	K_{MN}=\partial_M\partial_N B
\end{equation}
It is worth mentioning that the choice of the normal derivatives (also known as subsidiary conditions \cite{Bhattacharyya:2015dva, Bhattacharyya:2015fdk, Dandekar:2016fvw}) do not change the physical content of the corrected metrics at subsequent orders in $1/D$. The logic is based on the fact that the subsidiary conditions change the sub-leading piece of the ansatz metric and the metric corrections change accordingly to account for that. This is because the final metric with particular boundary conditions is unique. Hence, the physical content of the constraint gravity equation at $\psi=1$ which carries the  information of the membrane equations also are independent of the choice of subsidiary conditions (e.g. for two derivative theories see \cite{Bhattacharyya:2015fdk, Dandekar:2016fvw, Bhattacharyya:2017hpj}. 

Next we list together the objects that can contribute to the membrane equation which is obtained by evaluating the possible objects that can contribute to the constraint gravity equations on the $\psi=1$ slice and using our choice of subsidiary conditions. 

\subsection{Objects in the leading order membrane equations}
In this section we will list together all objects that can contribute to the scalar and vector membrane equations at leading order in $1/D$. Using the subsidiary conditions it is easy to see that $\mathcal{Q}$ evaluates to the trace of the local extrinsic curvature $\mathcal{K}$ of the $\psi=1$ membrane surface propagating in flat space-time. We consider two objects contributing to the membrane equation and differing only by a factor of $\mathcal{K}$ as equivalent. And hence in the gravity equations we did not distinguish objects differing by overall $\mathcal{Q}$ dependent factors. 

We have relegated the details of the procedure to appendix \eqref{gravtomemeqn}. The objects that can contribute to the scalar membrane equation at leading order are
\begin{equation}
	\boxed{1) u\cdot\hnabla\mathcal{K} \quad 2) u\cdot K\cdot u \quad 3) \hnabla^2\mathcal{K}}
\end{equation}
In the above $\hnabla$ denotes the covariant derivative w.r.t the induced metric in the world-volume of the membrane and $u$ is the velocity vector confined to the world-volume of the membrane. In addition we also list the possible objects that can contribute to the vector membrane equation at leading order. These objects are
\begin{eqnarray}
	&&\boxed{1)\left( \hnabla^2 u\cdot\mathcal{P}\right)_\mu, \quad 2) \left(\mathcal{P}\cdot\hnabla\mathcal{K}\right)_\mu\quad 3) \left(u\cdot\hnabla u\cdot \mathcal{P}\right)_\mu, \quad 4) \left(u\cdot K\cdot \mathcal{P}\right)_\mu}
\end{eqnarray}
where, $\mathcal{P}_{\mu\nu}=g_{\mu\nu}+u_\mu u_\nu$ and $g_{\mu\nu}$ is the induced metric on the membrane world volume. We write down some useful results now concerning membrane objects obtained from objects in the gravity equation at leading order. The first result is the tensorial decomposition of $\hnabla_\mu u_\nu$ w.r.t $\mathcal{P}_{\mu\nu}$, given by
\begin{eqnarray}\label{gradu}
	\nabla_\mu u_\nu&=& \sigma_{\mu\nu}+\omega_{\mu\nu}+a_\nu u_\mu+\frac{\mathcal{P}_{\mu\nu}}{D-2}\hnabla\cdot u\nonumber\\
	\text{where,}\quad \sigma_{\mu\nu}&=&\mathcal{P}^\alpha_\mu \frac{\hnabla_\alpha u_\beta+\hnabla _\beta u_\alpha}{2} \mathcal{P}^\beta_\nu-\frac{\mathcal{P}_{\mu\nu}}{D-2}\hnabla\cdot u\nonumber\\
	\omega_{\mu\nu}&=&\mathcal{P}^\alpha_\mu \frac{\hnabla_\alpha u_\beta-\hnabla _\beta u_\alpha}{2} \mathcal{P}^\beta_\nu\nonumber\\
	a_\nu&=& u\cdot \hnabla u_\nu
\end{eqnarray}
The next result is similar decomposition of the extrinsic curvature $K_{\mu\nu}$ given by
\begin{eqnarray}
	K_{\mu\nu}&=& K^{(TT)}_{\mu\nu}+u_\mu \left(u\cdot K\cdot \mathcal{P}\right)_\nu+u_\nu \left(u\cdot K\cdot \mathcal{P}\right)_\mu+u_\mu u_\nu \left(u\cdot K\cdot u\right)+\frac{\mathcal{P}_{\mu\nu}}{D-2}\mathcal{K}\nonumber\\
	\text{where,}\quad K^{(TT)}_{\mu\nu}&=&\mathcal{P}^\alpha_\mu K_{\alpha\beta}\mathcal{P}^\beta_\nu-\frac{\mathcal{P}_{\mu\nu}}{D-2}\mathcal{K}\nonumber\\
	\mathcal{K}&=&\mathcal{P}^{\mu\nu}K_{\mu\nu}
\end{eqnarray}
Another interesting result are the expressions of the world-volume divergences of the shear and vorticity vectors given by
\begin{eqnarray}\label{divofsigmaandomega}
	&&\hnabla^\mu\sigma_{\mu\alpha}=\frac{1}{2}\Bigg(\hnabla^2u_\beta +\mathcal{K}(u\cdot K)_\beta \Bigg)\mathcal{P}^\beta_\alpha+\mathcal{O}(1)\nonumber\\
	&&\hnabla^\mu\omega_{\mu\alpha}=\frac{1}{2}\Bigg(\hnabla^2u_\beta -\mathcal{K}(u\cdot K)_\beta \Bigg)\mathcal{P}^\beta_\alpha+\mathcal{O}(1)
\end{eqnarray}
To obtain the above result we use the Gauss-Codacci equations on the membrane world-volume. 
\subsection{Objects contributing to the sub-leading order scalar membrane equation}
For our analysis of the second law from the membrane equations at sub-leading order we need to focus only on the scalar membrane equation at this order. Hence, we only list the terms that can contribute to the scalar membrane equation. We classify the objects contributing to scalar membrane equation into two different classes
\begin{itemize}
	\item Bi-linears in the objects contributing to the first-sub-leading order gravity equations
	\item Objects which are not bi-linear. These can also include the objects present in the leading order scalar membrane equation.
\end{itemize}
\subsubsection{Objects bi-linear in leading order membrane objects}
We note that in the list of objects that can contribute to the second sub-leading order in gravity equations we have all possible products of pairs of objects which can contribute to the first sub-leading order gravity equations. Hence, we need not do any new analysis for this part and simply list the scalar objects formed by the product of leading order membrane objects listed in the last section.  The scalar objects from product of tensor objects are
\begin{equation}
	\boxed{1) \sigma_{\mu\nu}\sigma^{\mu\nu}\quad 2)\omega_{\mu\nu}\omega^{\mu\nu}\quad  3) K_{\mu\nu}^{(TT)}K^{(TT)\mu\nu}\quad 4) \sigma_{\mu\nu}K^{(TT)\mu\nu} }\
\end{equation}
The objects from the product of vector objects are
\begin{eqnarray}
	&&\boxed{5)a_\mu a^\mu\quad 6)(u\cdot K\cdot\mathcal{P})_\mu(u\cdot K\cdot\mathcal{P})^\mu\quad 7) (u\cdot K\cdot\mathcal{P})^\mu a_\mu\quad 8) (\hnabla^2 u\cdot \mathcal{P})_\mu a^\mu}\nonumber\\&& \boxed{9) (\hnabla^2 u\cdot \mathcal{P})_\mu (u\cdot K\cdot\mathcal{P})^\mu\quad 10) (\mathcal{P}\cdot\hnabla\mathcal{K})_\mu a^\mu\quad 11) (\mathcal{P}\cdot\hnabla_\mu\mathcal{K})(u\cdot K\cdot\mathcal{P})^\mu}\nonumber\\
	&&\boxed{12) (\mathcal{P}\cdot\hnabla\mathcal{K})^2\quad 13)\hnabla^2 u\cdot\mathcal{P}\cdot \hnabla^2 u\quad 14)\hnabla^2 u\cdot\mathcal{P}\cdot \hnabla\mathcal{K} }
\end{eqnarray}
And finally product of scalars given by
\begin{eqnarray}
	&&\boxed{15)(u\cdot K\cdot u)^2\quad 16)  (u\cdot\hnabla\mathcal{K})^2\quad 17) (\hnabla^2\mathcal{K})^2\quad 18)(u\cdot K\cdot u )(u\cdot\hnabla\mathcal{K})}\nonumber\\
	&&\boxed{19) (u\cdot K\cdot u)\hnabla^2\mathcal{K}\quad 20)(u\cdot\hnabla\mathcal{K})( \hnabla^2\mathcal{K})}
\end{eqnarray}	
\subsubsection{Non-bilinear objects}
The algorithm to arrive at the scalar objects from rest of the objects contributing to the second sub-leading gravity equation is similar (though more tedious). We won't present the details here and directly list the objects here. 
\begin{eqnarray}
	&&\boxed{1) \hnabla^2\mathcal{K}\quad 2) \nabla^2(u\cdot\hnabla\mathcal{K})\quad 3) u\cdot \hnabla\mathcal{K}\quad 4) u\cdot\hnabla(u\cdot\hnabla\mathcal{K})\quad 5) u\cdot \hnabla(u\cdot K\cdot u)}\nonumber\\
	&&\boxed{6) u\cdot K\cdot u\quad 7)\hnabla^2\hnabla^2\mathcal{K}}
\end{eqnarray}
The $\hnabla$ derivatives do not commute in general. And hence, one may think that there are more independent objects that can be formed by rearranging the order of the derivatives in the objects listed in this section. But we need not bother about this as objects obtained this way differ by factors of internal curvatures multiplied with objects with lower number of derivative. Since, internal curvatures are related to the extrinsic curvature by the Gauss-Codacci relations, these give rise to no new objects. 

Now that we have an exhaustive list of independent objects that can contribute to the membrane equations, in the next section we start the analysis for leading order membrane entropy current.
\section{The Leading order membrane entropy current}
From the list of terms that can contribute to the leading order membrane equation, we can write down the most general scalar membrane equations at this order as
\begin{eqnarray}\label{lead_scal_mem}
	\hnabla\cdot u=a_1(\alpha_K) u\cdot K\cdot u+a_2(\alpha_K)\frac{u\cdot\hnabla\mathcal{K}}{\mathcal{K}}+a_3(\alpha_K)\frac{\hnabla^2\mathcal{K}}{\mathcal{K}^2} +a_4(\alpha_K)\frac{\mathcal{K}}{D}+\mathcal{O}(1/D)\nonumber\\
\end{eqnarray}
where $a_1(\alpha_K)\ldots a_4(\alpha_K)$ are transport coefficients. Note that as mentioned earlier, we have multiplied the objects with appropriate powers of $\mathcal{K}$ and $D$ to make sure that the length dimensionality and orders in $D$ match on both sides. For the sake of generality we have included a term with $\mathcal{K}$ as an object. As we have argued earlier, the entropy current  is of the form
\begin{eqnarray}
	&&J_S^\mu=\frac{u^\mu}{4}+\mathcal{F}^\mu(\alpha_K)\nonumber\\
	\text{such that}\quad && \lim_{\alpha_K\rightarrow 0}\mathcal{F}^\mu(\alpha_K)=0
\end{eqnarray}
To arrive at a consistent entropy current we need to add appropriate counter-terms into $\mathcal{F}^\mu$ as the general membrane equation does not have a positive definite $\hnabla\cdot u$ at this order. Also, the entropy  current has to be a quantity in the membrane world volume. Hence we need
$$\boxed{a_4(\alpha_K)=0},$$
since, there is no world-volume vector quantity whose divergence gives $\mathcal{K}$. The divergence of the entropy current using the scalar membrane equation is given by
\begin{eqnarray}
	\hnabla\cdot J_S=\frac{a_1(\alpha_K)}{4}u\cdot K\cdot u+\frac{a_2(\alpha_K)}{4}\frac{u\cdot\hnabla\mathcal{K}}{\mathcal{K}}+a_3(\alpha_K)\frac{\hnabla^2\mathcal{K}}{4\mathcal{K}^2} +\hnabla\cdot \mathcal{F}
\end{eqnarray}
 The minimal choice of counter-terms that needs to be added to $\mathcal{F}^\mu$ to get rid of the non positive-definite terms above is
\begin{eqnarray}
	\mathcal{F}^\mu=-a_1(\alpha_K)\frac{(u\cdot\hnabla u)^\mu}{4\mathcal{K}}-a_2(\alpha_K)\frac{(u\cdot K)^\mu}{4\mathcal{K}}-a_3(\alpha_K)\frac{\hnabla^\mu\mathcal{K}}{4 \mathcal{K}^2}\nonumber\\
\end{eqnarray}
To show that the above counter-term takes care of the problematic terms in the divergence of entropy current, one needs to use the Guass-Codacci relations 
\begin{eqnarray}
	&&\hnabla^\mu K_{\mu\nu}=\hnabla_\nu \mathcal{K}\nonumber\\
	&& R_{\alpha\beta\gamma \delta}=K_{\alpha\gamma}K_{\beta\delta}-K_{\alpha\delta}K_{\beta\gamma}
\end{eqnarray}
where $R_{\alpha\beta\gamma\delta}$ is the intrinsic curvature of the induced metric on the membrane world-volume. And hence for the leading order scalar membrane equation \eqref{lead_scal_mem}, the leading order membrane entropy current with non-negative divergence is given by
\begin{eqnarray}
	J_S^{\mu(1)}&=&\frac{u^\mu}{4}-a_1(\alpha_K)\frac{(u\cdot\hnabla u)^\mu}{4\mathcal{K}}-a_2(\alpha_K)\frac{(u\cdot K)^\mu}{4\mathcal{K}}-a_3(\alpha_K)\frac{\hnabla^\mu\mathcal{K}}{4 D\mathcal{K}^2}\nonumber\\
	\hnabla\cdot J_S^{(1)}&=&\mathcal{O}(1/D)
\end{eqnarray}
i.e. for all classes of gravity theories that we are considering, entropy production starts at sub-leading order membrane equations. To find the entropy current whose divergence is positive definite, we have used the fact that the divergence of objects is usually at a higher order in $D$ than non-divergence terms. 
\subsection*{Leading order vector membrane equation}
The most general vector membrane equation is a linear combination of all possible membrane world volume vectors that we listed earlier. This can be written as
\begin{equation}\label{lead_mem_vec}
	\left(u\cdot\hnabla u_\mu+g_1(\alpha_K)\frac{\hnabla_\mu\mathcal{K}}{\mathcal{K}}+g_2(\alpha_K)\frac{\nabla^\nu \sigma_{\nu\mu}}{\mathcal{K}}+g_3(\alpha_K)\frac{\hnabla^\nu\omega_{\nu\mu}}{\mathcal{K}}\right)\mathcal{P}^\mu_\alpha=\mathcal{O}(1/D)
\end{equation}
Here we have used the pair of objects $\{\hnabla\cdot\sigma,\hnabla\cdot \omega\}$ instead of $\{\nabla^2u\cdot\mathcal{P},u\cdot K\cdot P\}$. We will find this choice convenient for the analysis of the next section. 
\subsection{Constraint from existence of stationary solutions at leading order}
In this section we impose constraint on the membrane equations from the existence of membrane configurations dual to stationary black hole. In \cite{Dandekar:2017aiv} it was shown that the membrane configurations dual to stationary black holes have a time-like killing vector field in the membrane world volume. This killing vector field is inherited by the membrane from the ambient space-time in which it is propagating. The membrane configuration is stationary if the Lie derivative of the membrane shape and velocity fields vanish along the time like killing vector field in the membrane world volume. 

Let the killing vector field be given by $k=k^\mu\partial_\mu.$ Then for stationarity we need
\begin{equation}
	\mathcal{L}_{k}u^\mu=k\cdot\hnabla u^\mu-u\cdot\hnabla k^\mu=0
\end{equation}
A simple velocity configuration which satisfies the above equation is if it is proportional to the killing vector field itself, i.e. 
\begin{equation}
	u^\mu=\gamma k^\mu\quad \text{where,}\quad \gamma=\frac{1}{\sqrt{-k\cdot k}}
\end{equation}
These configurations are found to be dual to stationary black holes in \cite{Mandlik:2018wnw} and we expect this to be true even here. 
Both the shear of the velocity field ($\sigma_{\mu\nu}$) and its divergence are zero for this stationary configuration. 
\begin{eqnarray}
	&&\hnabla\cdot u=\hnabla\cdot(\gamma k)=\gamma\hnabla\cdot k+k\cdot\hnabla\gamma=0\quad (\because \hnabla_{(\mu}k_{\nu)}=0, k\cdot \hnabla\gamma=0)\nonumber\\
	&&\sigma_{\mu\nu}=\mathcal{P}_\mu^\alpha\frac{\hnabla_\alpha (\gamma k_\beta)+\hnabla_\beta (\gamma k_\alpha)}{2}\mathcal{P}^\beta_\nu=\gamma \mathcal{P}_\mu^\alpha\frac{\hnabla_\alpha  k_\beta+\hnabla_\beta  k_\alpha}{2}\mathcal{P}^\beta_\nu=0\quad (\because k_\alpha \mathcal{P}^\alpha_\beta=0)\nonumber\\
\end{eqnarray}
But this velocity configuration has non-trivial acceleration and vorticity given by
\begin{eqnarray}
	a_\mu&=&u\cdot\nabla u_\mu=-\frac{\nabla_\mu\gamma}{\gamma}\nonumber\\
	\omega_{\mu\alpha}&=&\mathcal{P}^\nu_\mu\mathcal{P}^\beta_\alpha\frac{\nabla_\nu u_\beta-\nabla_\beta u_\nu}{2}=\gamma \mathcal{P}^\nu_\mu\mathcal{P}^\beta_\alpha\frac{\nabla_\nu k_\beta-\nabla_\beta k_\nu}{2}\nonumber\\
	\text{and,} \quad \nabla^\mu\omega_{\alpha\mu}\mathcal{P}^\alpha_\nu&=&-\gamma \mathcal{P}_\nu^\beta\nabla^\alpha(\nabla_\beta k_\alpha)+\mathcal{O}(D^0)=-\gamma \mathcal{P}_\theta^\beta\mathcal{K}\left(k^\delta K_{\delta\beta}\right)+\mathcal{O}(D^0)\nonumber\\
\end{eqnarray}
Where in the last line we have used the second Gauss-Codacci relation. Hence the vector membrane equation for this stationary configuration becomes
\begin{equation}
	\left(g_1(\alpha_K)\frac{\nabla_\mu\mathcal{K}}{\mathcal{K}}-\frac{\nabla_\mu\gamma}{\gamma}-g_3(\alpha_K)\gamma k^\beta K_{\beta\mu}\right)\mathcal{P}^\mu_\alpha=0
\end{equation}
Which can be re-written as
\begin{eqnarray}\label{stat_vec_eqn}
	&&\Bigg(\partial_\mu \Big(F(\alpha_K)-\ln(\gamma)\Big)-g_3(\alpha_K)\gamma k^\beta K_{\beta\mu}\Bigg)\mathcal{P}^\mu_\alpha=0\nonumber\\
	&&\text{where,}\quad 	F(\alpha_K)=\int\frac{g_1(\alpha_K)}{\mathcal{K}}d\mathcal{K}
\end{eqnarray}
There are $D-1$ membrane equations ($D-2$ vector equations and one scalar equations) for $D-1$ membrane variables ( $D-2$ independent components for unit normalised velocity field and one scalar function determining shape) in $D$ ambient space-time dimensions.~Given a initial membrane configuration where the velocity and shape function are known at all `spatial' points for a given initial constant time slice, the membrane equation gives as an output the velocity and shape configuration at all subsequent instances of time \footnote{There may be global/topological  constraints on the type of initial data allowed, but the membrane equations correctly capture the local evolution of the membrane data}. 

For the stationary configuration under consideration, the velocity field is given to be proportional to the killing vector field at all instances of time. Hence in a suitable coordinate system $D-2$ of the membrane equations are trivially solved and the remaining equation correctly captures the dynamics of the scalar membrane shape data. In a general coordinate system this amounts to all the different equations in \eqref{stat_vec_eqn} being equivalent to a single equation determining the dynamics of a scalar membrane shape data. For this to be true for \eqref{stat_vec_eqn} we need to have
\begin{equation}\label{vect_mem_eqn_constraint}
	\boxed{g_{3}(\alpha_K)=0}
\end{equation}
And the shape function for the stationary configurations with membrane velocity proportional to killing vector field is given by
\begin{equation}\label{stationary configuration}
\boxed{F(\alpha_K)-\ln(\gamma)=\text{constant}}
\end{equation}
The above equation should be interpreted as an algebraic equation on $\mathcal{K}$ for a given $k^\mu$ \footnote{This algebraic equation is a second order differential equation on the scalar function determining the local shape of the membrane}.

Both the constraint \eqref{vect_mem_eqn_constraint} as well as the stationary membrane equation above have been found to be true for all known examples in two derivative gravity \cite{Bhattacharyya:2016nhn, Mandlik:2018wnw} as well as for four derivative theories of gravity \cite{Saha:2018elg, Kar:2019kyz, Dandekar:2019hyc}. In fact the constraint \eqref{vect_mem_eqn_constraint} is also true for the charged membrane equations. 

Once we have constraint the vector membrane equations we can take a divergence of the vector membrane equation to to obtain
\begin{eqnarray}\label{div_of_vec_eqn}
	\boxed{\frac{\hnabla^2\mathcal{K}}{\mathcal{K}^2}=-\frac{1}{g_1(\alpha_K)}u\cdot K\cdot u-2\frac{g_2(\alpha_K)}{g_1(\alpha_K)}\frac{u\cdot\hnabla\mathcal{K}}{\mathcal{K}}+\mathcal{O}(D^0)}
\end{eqnarray}
Also, at the stationary configuration, taking the divergence of the acceleration and using the Gauss-Codacci equation we get
$$\mathcal{K}u\cdot K\cdot u=-\frac{\hnabla^2\gamma}{\gamma}+\mathcal{O}(D^0).$$
In addition by definition of stationary configuration $u\cdot\hnabla\mathcal{K}=0$.
Hence for the scalar membrane equation to make sense at stationary configuration we need  
 \begin{eqnarray}
\boxed{ a_1(\alpha_K)=\frac{a_3(\alpha_K)}{g_1(\alpha_K)}}
 \end{eqnarray}
\subsection{Summary of results for leading order membrane equations}
To conclude the analysis of the leading order membrane equations the most general scalar and vector membrane equations are given by
\begin{eqnarray}
	&&\boxed{\nabla\cdot u=A_2(\alpha_K)\frac{u\cdot\hnabla\mathcal{K}}{\mathcal{K}}+\mathcal{O}(1/D)\quad \text{where,}\quad A_2(\alpha_K)=a_2(\alpha_K)-2\frac{a_2(\alpha_K)g_2(\alpha_K)}{g_1(\alpha_K)}}\nonumber\\
	&&\boxed{\Big(u\cdot\hnabla u_\mu+g_1(\alpha_K)\frac{\hnabla_\mu\mathcal{K}}{\mathcal{K}}+g_2(\alpha_K)\frac{\hnabla^\nu\sigma_{\nu\mu}}{\mathcal{K}}\Big)\mathcal{P}^\mu_\alpha=\mathcal{O}(1/D)}
\end{eqnarray}	
The entropy current is given by
\begin{equation}
	J_S^{\mu(1)}=\frac{u^\mu}{4}-A_2(\alpha_K)\frac{(u\cdot K)^\mu}{4\mathcal{K}}
\end{equation}
Taking the divergence of the above entropy current and using the membrane equations we obtain
\begin{equation}
	\hnabla\cdot J_S^{(1)}=\mathcal{O}(1/D)
\end{equation}
 Presence of any other structure in the membrane equations is not consistent with the presence of stationary membrane configurations or with consistent entropy current.
\section{List of terms in scalar equation and possible counter-terms in $J_S$}
Before analysing the contribution to the entropy current for the sub-leading order membrane equations we write down the contribution at $\mathcal{O}(1/D)$ of the divergence of the leading order contributions to the entropy current
\begin{eqnarray}
	\hnabla\cdot J_S^{(1)}&=& -\partial_\mathcal{K}\left(\frac{A_2(\alpha_K)}{4\mathcal{K}}\right)\partial_\mu\mathcal{K} (u\cdot K)^\mu-\frac{A_2(\alpha_K)}{4\mathcal{K}}\hnabla_\mu u_\nu K^{\mu\nu}+\mathcal{O}(1/D^2)\nonumber\\
	&=&\partial_\mathcal{K}\left(\frac{A_2(\alpha_K)}{4\mathcal{K}}\right)\Big(u\cdot\hnabla\mathcal{K}u\cdot K\cdot u-\partial_\mu\mathcal{K} u\cdot\hnabla u_\nu \mathcal{P}^\mu\nu\Big)\nonumber\\
	&&-\frac{A_2(\alpha_K)}{4\mathcal{K}}\Big(\sigma_{\mu\nu}K^{\mu\nu(TT)}+u\cdot\hnabla u_\mu (u\cdot K)_\nu \mathcal{P}^{\mu\nu}\Big)-\frac{(A_2(\alpha_K))^2}{4 D\mathcal{K}}u\cdot\hnabla\mathcal{K}+\mathcal{O}(1/D^2)\nonumber\\
\end{eqnarray}
 The above expression contains objects which also appear in the scalar membrane equation at sub-leading order and hence we have to be careful about the coefficients of these terms while writing down the entropy current at sub-leading order. 
The general structure of the entropy current is of the form. 
\begin{eqnarray}
	J_S^\mu=\frac{u^\mu}{4}+J_S^{(1)\mu}+J_S^{(2)\mu}
\end{eqnarray}
$J^{(1)}_S$ and $J_S^{(2)}$ are respectively the contributions to the entropy current due to the leading and sub-leading order membrane equations. So, 
\begin{eqnarray}
	\hnabla\cdot J_S=\frac{\hnabla\cdot u}{4}+\hnabla\cdot J^{(1)}_S+\hnabla\cdot J_S^{(2)}
\end{eqnarray}

The $\mathcal{O}(1/D)$ terms in the the $\nabla\cdot J_S^{(1)}$ can effectively be taken into account shifting these terms to the definition of $\hnabla\cdot u$ at sub-leading order. Hence, we can take into account the sub-leading order effects of objects appearing in the leading order membrane equation by adding the following extra contribution to the $\hnabla\cdot u$

\begin{eqnarray}\label{effec_div_u}
	\hnabla\cdot u&=&\partial_\mathcal{K}\left(\frac{A_2(\alpha_K)}{\mathcal{K}}\right)\Big(u\cdot\hnabla\mathcal{K}u\cdot K\cdot u-\partial_\mu\mathcal{K} u\cdot\hnabla u_\nu \mathcal{P}^\mu\nu\Big)\nonumber\\
	&&-\frac{A_2(\alpha_K)}{\mathcal{K}}\Big(\sigma_{\mu\nu}K^{\mu\nu(TT)}+u\cdot\hnabla u_\mu u\cdot K_\nu \mathcal{P}^{\mu\nu}\Big)-\frac{(A_2(\alpha_K))^2}{ D\mathcal{K}}u\cdot\hnabla\mathcal{K}+\mathcal{O}(1/D^2)\nonumber\\
\end{eqnarray}
which using the leading order vector membrane equation can be written as
\begin{eqnarray}	
	\hnabla\cdot u&=&\partial_\mathcal{K}\left(\frac{A_2(\alpha_K)}{\mathcal{K}}\right)\Big(u\cdot\hnabla\mathcal{K}u\cdot K\cdot u+\frac{\mathcal{K}}{g_1(\alpha_K)}u\cdot\hnabla u_\mu u\cdot\hnabla u^\mu\nonumber\\
	&&+\frac{g_2(\alpha_K)}{2g_1(\alpha_K)}u\cdot\hnabla u^\mu\left(\hnabla^2 u_\mu+\mathcal{K}(u\cdot K)_\mu\right)\Big)-\frac{A_2(\alpha_K)}{\mathcal{K}}\Big(\sigma_{\mu\nu}K^{\mu\nu(TT)}+u\cdot\hnabla u_\mu u\cdot K_\nu \mathcal{P}^{\mu\nu}\Big)\nonumber\\
	&&-\frac{(A_2(\alpha_K))^2}{ D\mathcal{K}}u\cdot\hnabla\mathcal{K}+\mathcal{O}(1/D^2)
\end{eqnarray}
Also, for the analysis of the sub-leading order membrane equation and entropy current we assume that the leading order membrane equations are solved upto relevant order. This we can ensure by demanding that $\nabla_\mu \mathcal{K}$ and $\nabla^2\mathcal{K}$ are determined in terms of other vectors and scalars using the vector equation and its divergence. Hence, we remove this objects, their derivatives or bi-linears involving them from the list of possible objects in the membrane equation at sub-leading order. 
\subsection{Contributions from objects also present in leading order membrane equation}
The contributions due to these objects to the sub-leading order membrane equations are given by
\begin{eqnarray}
	\hnabla\cdot u=c_1(\alpha_k)\frac{u\cdot K\cdot u}{D}+c_2(\alpha_K)\frac{u\cdot\hnabla\mathcal{K}}{D\mathcal{K}}
\end{eqnarray} 
Taking into account the $\mathcal{O}(1/D)$ effects due to leading order entropy current (see \eqref{effec_div_u}), the coefficients above change to
\begin{eqnarray}
	\hnabla\cdot u=c_1(\alpha_k)\frac{u\cdot K\cdot u}{D}+\Big(c_2(\alpha_K)-(A_2(\alpha_K))^2\Big)\frac{u\cdot\hnabla\mathcal{K}}{D\mathcal{K}}
\end{eqnarray}
The above contributions to $\hnabla\cdot u$ are not positive definite and the counter-term corresponding to the above two terms is given by
\begin{eqnarray}
	\mathcal{F}^\mu_2(\alpha_K)=-c_1(\alpha_K)\frac{(u\cdot\hnabla u)^\mu}{4\mathcal{K}D}-\Big(c_2(\alpha_K)-(A_2(\alpha_K))^2\Big)\frac{(u\cdot K)^\mu}{4\mathcal{K}D}
\end{eqnarray}
\subsection{New non-bilinear objects at sub-leading order}
There are three new non bilinear objects which contribute to the sub-leading order scalar membrane equation
\begin{eqnarray}
	\hnabla\cdot u=b_1(\alpha_K)\frac{u\cdot \nabla(u\cdot K\cdot u)}{\mathcal{K}} +b_2(\alpha_K)\frac{u\cdot \nabla(u\cdot \nabla \mathcal{K})}{\mathcal{K}^2}+b_3(\alpha_K)\frac{\nabla^2(u\cdot \nabla \mathcal{K})}{\mathcal{K}^3}
\end{eqnarray}
The counter-term necessary for the above contributions is given by
\begin{eqnarray}
	\mathcal{F}_2^\mu=-b_1(\alpha_K)\frac{u\cdot \nabla(u\cdot \nabla u)^\mu}{4\mathcal{K}^2}-b_2(\alpha_K)\frac{u\cdot \nabla(u\cdot K)^\mu}{\mathcal{K}^2}-b_3(\alpha_K)\frac{\nabla^\mu(u\cdot\nabla\mathcal{K})}{4\mathcal{K}^3}
\end{eqnarray}
\subsection{Bilinear objects}
Next we analyse the contributions of the bi-linear terms one by one
\subsubsection*{$1)\underline{\bf{(u\cdot\hnabla\mathcal{K})^2}}$}
\begin{equation}
	\hnabla\cdot u= d_1(\alpha_K)\frac{(u\cdot \nabla \mathcal{K})^2}{\mathcal{K}^3}
\end{equation}
The counter-term is given by
\begin{equation}
	\mathcal{F}^\mu=-d_1(\alpha_K)\frac{(u\cdot K)^\mu u\cdot \nabla\mathcal{K}}{4\mathcal{K}^3}
\end{equation}
One thing to note here is that we are considering systems with $SO(D-p-3)$ isometry where $p$ is held fixed and $D\rightarrow\infty$. In this situation $\mathcal{K}>0$ and hence the above contribution to the membrane equation is such that if we had imposed the constraint 
$d_1(\alpha_K)\ge 0,$
then we could have gotten away without adding a counter-term to the entropy current for this object. But we follow the principle of imposing the least number of constraints on the ``transport coefficients" and then adding the minimal number of terms to the entropy current. And hence we chose to add the counter-term for this objects. 
\subsubsection*{$2)\underline{\bf{(u\cdot K\cdot u)^2}}$}
\begin{equation}
	\hnabla\cdot u=d_2(\alpha_K)\frac{(u\cdot K\cdot u)^2}{\mathcal{K}}
\end{equation}
The counter-term for the above contribution is
\begin{equation}
	\mathcal{F}^\mu=-d_2(\alpha_K)\frac{(u\cdot\nabla u)^\mu(u\cdot K\cdot u)}{4\mathcal{K}^2}
\end{equation}
\subsubsection*{$3)\underline{\bf{(u\cdot K\cdot u)(u\cdot\hnabla\mathcal{K})}}$}
\begin{equation}
	\hnabla\cdot u= d_3(\alpha_K)\frac{(u\cdot\nabla\mathcal{K})(u\cdot K\cdot u)}{\mathcal{K}^2}
\end{equation}
Taking into account the $\mathcal{O}(1/D)$ effect of the leading order entropy current, the contribution of the above term gets modified to
\begin{equation}
	\hnabla\cdot u= \Big(d_3(\alpha_K)+\mathcal{K}\partial_\mathcal{K}A_2(\alpha_K)-A_2(\alpha_K)\Big)\frac{(u\cdot\nabla\mathcal{K})(u\cdot K\cdot u)}{\mathcal{K}^2}
\end{equation}
There are two possible  counter-term for the above term which are
\begin{eqnarray}
	&&\mathcal{F}^\mu=-\bar{d}_3(\alpha_K)\frac{(u\cdot K)^\mu(u\cdot K\cdot u)}{4\mathcal{K}^2}\quad \text{or,}\quad -\bar{d}_3(\alpha_K)\frac{(u\cdot \nabla \mathcal{K})(u\cdot\nabla u)^\mu}{4\mathcal{K}^3}\nonumber\\
	&&\text{where,}\quad  \bar{d}_3(\alpha_K)=\Big(d_3(\alpha_K)+\mathcal{K}\partial_\mathcal{K}A_2(\alpha_K)-A_2(\alpha_K)\Big)
\end{eqnarray}
\subsubsection*{$4)\underline{\bf{\hnabla^2u_\alpha\hnabla^2 u_\beta \mathcal{P}^{\alpha\beta}}}$}
\begin{equation}
	\nabla\cdot u=d_4(\alpha_K)\frac{\nabla^2 u_\mu\nabla^2 u_\nu\mathcal{P}^{\mu\nu}}{\mathcal{K}^3}
\end{equation}
The counter-term is given by
\begin{equation}
	\mathcal{F}^\mu=-d_4(\alpha_K)\frac{\nabla^\mu u_\alpha \nabla^2 u_\beta \mathcal{P}^{\alpha\beta}}{4\mathcal{K}^3}
\end{equation}
\subsubsection*{$5)\underline{\bf{u\cdot\hnabla u_\alpha u\cdot\hnabla u_\beta \mathcal{P}^{\alpha\beta}}}$}
\begin{equation}
	\hnabla\cdot u= d_5(\alpha_K)\frac{(u\cdot \nabla u)_\mu (u\cdot\nabla u)_\nu\mathcal{P}^{\mu\nu}}{\mathcal{K}}
\end{equation}
Again taking into account sub-leading order effect of leading order entropy current, the modified contribution of this object is given by
\begin{equation}
	\hnabla\cdot u=\bar{d}_5(\alpha_K)\frac{(u\cdot \nabla u)_\mu (u\cdot\nabla u)_\nu\mathcal{P}^{\mu\nu}}{\mathcal{K}}, \quad \text{where,}\quad \bar{d}_5(\alpha_K)=d_5(\alpha_K)+\frac{\mathcal{K}\partial_{\mathcal{K}} A_2(\alpha_K)-A_2(\alpha_K)}{g_1(\alpha_K)}
\end{equation}
There is no counter-term corresponding to this object. Since, the norm of the acceleration is positive definite, this contribution to the membrane equation will give rise to a positive definite contribution to the divergence of entropy current provided we have
\begin{equation}
	\boxed{\bar{d}_5(\alpha_K)\ge0}
\end{equation}
\subsubsection*{$6)\underline{\bf{(u\cdot K\cdot\mathcal{P})_\mu(u\cdot K\cdot \mathcal{P})^\mu}}$}
\begin{equation}
	\hnabla\cdot u=d_6(\alpha_K)\frac{(u\cdot K)_\nu(u\cdot K)_\nu\mathcal{P}^{\mu\nu}}{\mathcal{K}}
\end{equation}
The corresponding counter-term is given by
\begin{equation}
	\mathcal{F}^\mu=-d_6(\alpha_K)\frac{\hnabla_\nu u^\mu(u\cdot K)_\alpha\mathcal{P}^{\alpha\nu}}{4\mathcal{K}^2}
\end{equation}
\subsubsection*{$7)\underline{\bf{(u\cdot K\cdot \mathcal{P})^\mu u\cdot\hnabla u_\mu}}$}
\begin{equation}
	\nabla\cdot u=d_7(\alpha_K)\frac{(u\cdot K)_\mu(u\cdot\hnabla u)_\nu\mathcal{P}^{\mu\nu}}{\mathcal{K}}
\end{equation}
Taking into account the sub-leading order effect of leading order entropy current the above equation is modified to
\begin{eqnarray}
	&&\nabla\cdot u=\bar{d}_7(\alpha_K)\frac{(u\cdot K)_\mu(u\cdot\hnabla u)_\nu\mathcal{P}^{\mu\nu}}{\mathcal{K}}\nonumber\\
	&&\text{where,}\quad \bar{d}_7(\alpha_K)=d_7(\alpha_K)+\frac{g_2(\alpha_K)}{g_1(\alpha_K)}\Bigg(\mathcal{K}\partial_{\mathcal{K}}A_2(\alpha_K)-A_2(\alpha_K)\left(1+\frac{2g_1(\alpha_K)}{g_2(\alpha_K)}\right)\Bigg)\nonumber\\
\end{eqnarray}
the counter-terms is
\begin{equation}
	\mathcal{F}^\mu=-\bar{d}_7(\alpha_K)\frac{\hnabla_\nu u^\mu(u\cdot\hnabla u)_\alpha\mathcal{P}^{\alpha\nu}}{4\mathcal{K}^2}
\end{equation}
\subsubsection*{$8)\underline{\bf{\hnabla^2 u_\alpha u\cdot\hnabla u_\beta \mathcal{P}^{\alpha\beta}}}$}
\begin{equation}
	\hnabla\cdot u=d_8(\alpha_K) \frac{(u\cdot\hnabla u)_\nu\hnabla^2 u_\beta \mathcal{P}^{\nu\beta}}{\mathcal{K}^2}
\end{equation}
Once again the coefficient of this term is modified by the sub-leading effect of leading order entropy current and is given by
\begin{equation}
	\hnabla\cdot u=\bar{d}_8(\alpha_K) \frac{(u\cdot\hnabla u)_\nu\hnabla^2 u_\beta \mathcal{P}^{\nu\beta}}{\mathcal{K}^2}\quad \text{where,}\quad \bar{d}_8(\alpha_K)=d_8(\alpha_K)+\frac{g_2(\alpha_K)}{g_1(\alpha_K)}\Big(\mathcal{K}\partial_{\mathcal{K}}A_2(\alpha_K)-A_2(\alpha_K)\Big)
\end{equation}
The counter-term is given by
\begin{equation}
	\mathcal{F}^\mu=-\bar{d}_8(\alpha_K)\frac{(u\cdot\hnabla u)_\nu \hnabla^\mu u_\beta \mathcal{P}^{\nu\beta}}{4\mathcal{K}^2}
\end{equation}
\subsubsection*{$9)\underline{\bf{\hnabla^2 u_\alpha (u\cdot K)_\beta \mathcal{P}^{\alpha\beta}}}$}
The contribution to the scalar membrane equation is given by
\begin{eqnarray}
	\hnabla\cdot u=d_9(\alpha_K)\frac{\hnabla^2 u_\alpha (u\cdot K)_\beta \mathcal{P}^{\alpha\beta}}{\mathcal{K}^2}
\end{eqnarray}
Again there are two possible counter-terms given by
\begin{eqnarray}
	\mathcal{F}^\mu=-d_9(\alpha_K)\frac{\hnabla^\mu u_\alpha (u\cdot K)_\beta \mathcal{P}^{\alpha\beta}}{4\mathcal{K}^2}\quad \text{or,}\quad \mathcal{F}^\mu=-d_9(\alpha_K)\frac{(\hnabla^2 u\cdot \mathcal{P})_\alpha\hnabla^\alpha u^\mu}{4\mathcal{K}^3}
\end{eqnarray}
\subsubsection*{$10)\underline{\bf{\sigma_{\alpha\beta}\sigma^{\alpha\beta}}}$}
\begin{equation}
	\hnabla\cdot u=d_{10}(\alpha_K)\frac{\sigma_{\alpha\beta}\sigma^{\alpha\beta}}{\mathcal{K}}
\end{equation}
The normal of shear tensor is positive definite and hence and in addition there exists no counter-term for the above term and only possible way for this to give rise to a positive definite contribution to the divergence of entropy current is if we impose
\begin{equation}
	\boxed{d_{10}(\alpha_K)\ge 0}
\end{equation}

\subsubsection*{$11)\underline{\bf{\omega_{\alpha\beta}\omega^{\alpha\beta}}}$}
\begin{equation}
	\hnabla\cdot u=d_{11}(\alpha_K)\frac{\omega_{\mu\alpha}\omega^{\mu\alpha}}{\mathcal{K}}
\end{equation}
This term also does not have a counter-term and again the norm of vorticity is positive definite. Hence for positive definite local entropy production we must again have
\begin{equation}
	\boxed{d_{11}(\alpha_K)\ge 0}
\end{equation}

\subsubsection*{$12)\underline{\bf{K^{(TT)}_{\alpha\beta}K^{(TT)\alpha\beta}}}$}
\begin{equation}
	\hnabla\cdot u=d_{12}(\alpha_K)\frac{K^{(TT)}_{\mu\alpha}K^{(TT)\mu\alpha}}{\mathcal{K}}
\end{equation}
The object $K^{(TT)}_{\mu\nu}$ captures the component of the extrinsic curvatures orthogonal to the velocity vector field and hence its norm is positive definite. Also, this object has no corresponding counter-term and will contribute to positive definite entropy production provided
\begin{equation}
	\boxed{d_{12}(\alpha_K)\ge0}
\end{equation}
The analysis of the last object that can possibly be present in the sub-leading scalar membrane equation is a bit involved and we dedicate the next sub-section for the analysis. 
\subsection{Analysis of the object $\sigma_{\alpha\beta}K^{(TT)\alpha\beta}$ in the scalar membrane equation}
This contribution of this object to the scalar membrane equation is special. This is clear from the derivation of the entropy current for Einstein-Gauss-Bonnet gravity perturbatively in GB parameter in \cite{Dandekar:2019hyc} which we reviewed in one of the earlier sections. For this analysis we need to consider the schematic form of the scalar membrane equation given by
\begin{equation}
	\nabla\cdot u=a(\alpha_K)+Q^2(\alpha_k)+ b(\alpha_K)\frac{\sigma_{\alpha\beta}K^{(TT)\alpha\beta}}{\mathcal{K}}.
\end{equation}
Here $a(\alpha_K)$ represents those terms in the scalar membrane equation which have a corresponding counter-term in the entropy current expression. The term $Q^2$ above denotes all terms in the scalar membrane equation which do not have a counter-term and which are positive definite provided their "transport coefficients" satisfy some constraints. Example of object in $a(\alpha_K)$ are $u\cdot\hnabla \mathcal{K}$ and $u\cdot\hnabla\cdot(\hnabla\mathcal{K})$ . Examples of objects in $Q^2(\alpha_K)$ are $\omega_{\mu\nu}\omega^{\mu\nu}$ and $a_\mu a^\mu$. 

We will assume that sub-leading effect of the leading order entropy current have been taken into account here. In particular 
\begin{equation}
	b(\alpha_K)=\tilde{b}(\alpha_K)-\frac{A_2(\alpha_K)}{\mathcal{K}}
\end{equation}
where, $\tilde{b}(\alpha_K)$ is the coefficient with which the object actually contributes to the sub-leading order scalar membrane equation. 

 From the perturbative analysis of EGB gravity, the following choice of entropy current naively looks like a good choice to get a positive definite entropy production from the above scalar membrane equation. 
\begin{eqnarray}
	&&J_S^\mu=\frac{u^\mu}{4}-j_a^\mu-\frac{b(\alpha_K)}{4}\frac{(u\cdot K)^\mu}{\mathcal{K}}+\mathcal{S}^\mu\nonumber\\
	&&\text{where,}\quad \hnabla\cdot j_a=\frac{a(\alpha_K)}{4}
\end{eqnarray}
We have added an extra counter-term $\mathcal{S}^\mu$ whose form we will choose later depending upon the non-positive definite objects that appear in $\hnabla\cdot J$. Taking the divergence of this entropy current and using the schematic scalar membrane equation gives 
\begin{equation}
	\nabla\cdot J_S=\frac{1}{4}Q^2(\alpha_K)-\frac{(u\cdot K)^\mu}{4}\left(\partial_{\mathcal{K}}\left(\frac{b(\alpha_K)}{\mathcal{K}}\right)\partial_\mu\mathcal{K}+b(\alpha_K)\frac{(u\cdot\nabla u)_\mu}{\mathcal{K}}\right)-b(\alpha_K)\frac{u\cdot\nabla\mathcal{K}}{4\mathcal{K}}+\nabla\cdot\mathcal{S}
\end{equation}
The second and third term on the RHS above are not positive definite and hence we add the following terms to $\mathcal{S}^\mu$
\begin{eqnarray}
	\mathcal{S}^\mu&=&W(\alpha_K)u^\mu+P^\mu\nonumber\\
	\text{where,}\quad W(\alpha_K)&=&\int\frac{b(\alpha_K)}{4\mathcal{K}}dK\nonumber\\
	\text{and,}\quad P^\mu&=&-\frac{\nabla_\nu u^\mu}{\mathcal{K}}\left(\partial_{\mathcal{K}}\left(\frac{b(\alpha_K)}{\mathcal{K}}\right)\partial_\mu\mathcal{K}+b(\alpha_K)\frac{(u\cdot\nabla u)_\mu}{\mathcal{K}}\right)
\end{eqnarray}
So, that we have
\begin{eqnarray}
	\hnabla\cdot J_S&=&\frac{1}{4}Q^2(\alpha_K)+W(\alpha_K)\nabla\cdot u\nonumber\\
	&=&\frac{1}{4}Q^2(\alpha_K)+W(\alpha_K)\left(a(\alpha_K)+Q^2(\alpha_k)+ b(\alpha_K)\frac{\sigma_{\alpha\beta}K^{(TT)\alpha\beta}}{\mathcal{K}}\right)\nonumber\\
	&=&\frac{1}{4}Q^2(\alpha_K)\Bigg(1+4W(\alpha_K)\Bigg)+W(\alpha_K)\left(a(\alpha_K)+ b(\alpha_K)\frac{\sigma_{\alpha\beta}K^{(TT)\alpha\beta}}{\mathcal{K}}\right)\nonumber\\
\end{eqnarray}
where, in the second line above we have used the schematic form of the scalar membrane equation. We see that the presence of the object $\sigma^{\alpha\beta}K_{\alpha\beta}^{(TT)}$ in scalar membrane equation forces us to add the counter-term proportional to $u^\mu$ above which in turn gives rise to the second term above. We see the divergence of the above entropy current is not positive definite and we also see that the object $\sigma_{\mu\nu}K^{(TT)\mu\nu}$ emerges back.

Nevertheless to understand the suggestive recurrent pattern above we do one more iteration and propose a modified entropy current given by
\begin{eqnarray}
	\tilde{J}_S^\mu&=&J_S^\mu-4W(\alpha_K)\left(j^\mu_a+b(\alpha_K)\frac{(u\cdot K)^\mu}{4\mathcal{K}}\right)+\tilde{S}^\mu\nonumber\\
	\text{where,}\quad J_S^\mu&=& \frac{u^\mu}{4}\Bigg(1+4W(\alpha_K)\Bigg)-j_a^\mu-b(\alpha_K)\frac{(u\cdot K)^\mu}{4\mathcal{K}}+P^\mu
\end{eqnarray}
So, that
\begin{eqnarray}
	\nabla\cdot\tilde{J}_S&=&\frac{1}{4}Q^2(\alpha_K)\Bigg(1+W(\alpha_K)\Bigg)-(u\cdot K)^\nu\Bigg(W(\alpha_K)b(\alpha_K)\frac{(u\cdot\nabla u)_\nu}{\mathcal{K}}+\partial_{\nu}\Bigg(\frac{W(\alpha_K)b(\alpha_K)}{\mathcal{K}}\Bigg)\Bigg)\nonumber\\&&-W(\alpha_K)b(\alpha_K)\frac{u\cdot\nabla\mathcal{K}}{\mathcal{K}}+\nabla\cdot\tilde{S}
\end{eqnarray}
To try to get rid of the non-positive definite second and third term above we again guess the counter-term $\tilde{S}^\mu$ to be given by
\begin{eqnarray}
	\tilde{S}^\mu&=&\tilde{W}(\alpha_K)u^\mu+\tilde{P}^\mu\quad \text{where,}\nonumber\\
  \tilde{W}(\alpha_K)&=&\int\frac{W(\alpha_K)b(\alpha_K)}{\mathcal{K}}dK= \frac{1}{4}\int\frac{b(\alpha_K)}{\mathcal{K}}\left(\int\frac{b(\alpha_K)}{\mathcal{K}}dK\right)dK\nonumber\\
	\tilde{P}^\mu&=&\frac{\nabla_\nu u^\mu}{\mathcal{K}}\Bigg(W(\alpha_K)b(\alpha_K)\frac{(u\cdot\nabla u)_\nu}{\mathcal{K}}+\partial_{\nu}\Bigg(\frac{W(\alpha_K)b(\alpha_K)}{\mathcal{K}}\Bigg)\Bigg)\nonumber\\
	&=&4W(\alpha_K) P^\mu+b(\alpha_K)\frac{\nabla_\nu u^\mu}{\mathcal{K}}\partial^\nu W(\alpha_K)
\end{eqnarray}
The divergence of the modified entropy current is given by
\begin{eqnarray}
	\nabla\cdot \tilde{J}_S&=&\frac{1}{4}Q^2(\alpha_K)\Bigg(1+4W(\alpha_K)+4\tilde{W}(\alpha_K)\Bigg)+\tilde{W}\Bigg(a(\alpha_K)+b(\alpha_K)\frac{\sigma_{\alpha\beta}K^{\alpha\beta}}{\mathcal{K}}\Bigg)\nonumber\\
	\text{where,}\nonumber\\ \tilde{J}^\mu_S&=&\frac{u^\mu}{4}\Bigg(1+4W(\alpha_K)+4\tilde{W}(\alpha_K)\Bigg)-\left(j^\mu_a+\frac{b(\alpha_k)}{4\mathcal{K}}(u\cdot K)^\mu-P^\mu\right)\Big(1+4W(\alpha_K)\Bigg)\nonumber\\
	&&b(\alpha_K)\frac{\nabla_\nu u^\mu}{\mathcal{K}}\partial^\nu W(\alpha_K)\nonumber\\
	\quad P^\mu&=&-\frac{\nabla_\nu u^\mu}{\mathcal{K}}\left(\partial_{\mathcal{K}}\left(\frac{b(\alpha_K)}{\mathcal{K}}\right)\partial_\mu\mathcal{K}+b(\alpha_K)\frac{(u\cdot\nabla u)_\mu}{\mathcal{K}}\right) \nonumber\\
	\tilde{W}(\alpha_K)&=&\frac{1}{4}\int\frac{b(\alpha_K)}{\mathcal{K}}\left(\int\frac{b(\alpha_K)}{\mathcal{K}}dK\right)dK\nonumber\\
	W(\alpha_K)&=&\frac{1}{4}\int\frac{b(\alpha_K)}{\mathcal{K}}dK
\end{eqnarray}
Therefore, we land in a similar position as earlier. From this two iterations we can guess the iterative structure of the entropy current and its divergence. e.g the coefficient of the term proportional to velocity vector in the entropy current after the third iteration will be given by
$$1+4W(\alpha_K)+4\tilde{W}(\alpha_K)+4\tilde{\tilde{W}}(\alpha_K),$$
where, $$\tilde{\tilde{W}}(\alpha_K)=\frac{1}{4}\int\frac{ b(\alpha_K)}{\mathcal{K}}\Bigg(\int\frac{b(\alpha_K)}{\mathcal{K}}\Bigg(\int\frac{b(\alpha_K)}{\mathcal{K}}dK\Bigg)dK\Bigg)dK$$
and the coefficient of $Q^2(\alpha_K)$ in the divergence of the entropy current will be given by
$$1+4W(\alpha_K)+4\tilde{W}(\alpha_K)+4\tilde{\tilde{W}}(\alpha_K).$$
A few comments about the consequence of presence of a non-zero transport coefficient $b(\alpha_K)$ are in order here. 
\begin{enumerate}
	\item Unless the series of integral 
	$$\frac{1}{4}\int\frac{ b(\alpha_K)}{\mathcal{K}}\Bigg(\int\frac{b(\alpha_K)}{\mathcal{K}}\Bigg(\int\frac{b(\alpha_K)}{\mathcal{K}}dK\Bigg)\ldots dK\Bigg)dK$$
	evaluates to zero, there is no counter-term which can give rise to a positive definite entropy production in presence of non-zero $b(\alpha_K)$. Hence, if we are working non-perturbatively in $\alpha_K$ the only way to get a positive definite entropy production is if we have
	$$\boxed{b(\alpha_K)=0}$$
	\item If we are working perturbatively in $\alpha_K$ upto some order, then the series of integral truncates to give an entropy current with positive definite entropy production upto the desired order in $\alpha_K$. This is what we observed in \cite{Dandekar:2019hyc} for a perturbative analysis of Einstein-Gauss-Bonnet gravity. Hence, the obstruction to second law from non-zero $b(\alpha_K)$ is purely non-perturbative. 
	\item The fact that naively a perturbative analysis seems to give a consistent result and non-perturbative analysis forces us to set $b(\alpha_K)=0$ hints towards the fact that the perturbative expression of entropy current in presence of non-zero $b(\alpha_K)$ may have zero radius of convergence in complex $\alpha_K$ plane. 
	\item $a(\alpha_K)$ in general can be $\mathcal{O}(D^0)$ and hence, presence $b(\alpha_K)$ can give rise to $\mathcal{O}(D^0)$ entropy production even though the leading order membrane equation gave rise to zero entropy production at $\mathcal{O}(D^0)$. Hence, for consistency of the $1/D$ perturbative method we require $$\boxed{b(\alpha_K)=0\quad \text{or,}\quad  A_2(\alpha_K)=0}$$
	For all known examples it is indeed true that $A_2(\alpha_K)=0.$
\end{enumerate}
\section{ Constraints  from existence of Stationary solution}
We now study the effect on various transport coefficients by demanding the presence of a stationary membrane configuration with
$$u^\mu=\gamma k^\mu.$$
The net divergence of entropy current at sub-leading order in $1/D$ at the stationary configuration is given by
\begin{equation}
	\hnabla\cdot J_S=d_5(\alpha_K)\frac{(u\cdot \nabla u)_\mu (u\cdot\nabla u)^\mu}{4\mathcal{K}}+d_{10}(\alpha_K)\frac{\sigma_{\alpha\beta}\sigma^{\alpha\beta}}{4\mathcal{K}}+d_{11}(\alpha_K)\frac{\omega_{\mu\alpha}\omega^{\mu\alpha}}{4\mathcal{K}}+d_{12}(\alpha_K)\frac{K^{(TT)}_{\mu\alpha}K^{(TT)\mu\alpha}}{4\mathcal{K}}
\end{equation}
For the stationary configuration under consideration here we know that the membrane configuration has vanishing shear, but non-vanishing acceleration and vorticity as well as non-zero $K_{\mu\nu}^{(TT)}$. Also a stationary configuration by definition is characterised by no net entropy production. Since, the entropy production is positive definite locally at each point on the membrane, for net zero entropy production we require that the divergence of the entropy current vanish for these configuration. This imposes the constraints 
\begin{equation}
	\boxed{d_5(\alpha_K)=0}\quad \boxed{d_{11}(\alpha_K)=0}\quad \text{and,}\quad \boxed{d_{12}(\alpha_K)=0}.
\end{equation}
For the stationary configuration $u^\mu=\gamma k^\mu$ we also have $$\hnabla\cdot u=0.$$ This imposes additional constraints on the transport coefficient of the scalar membrane equation. From our analysis of the leading in large $D$ we know that $$u\cdot K\cdot u=-\frac{\hnabla^2 \gamma}{\mathcal{K} \gamma}\ne 0$$
Hence we also, need to set
\begin{equation}
	\boxed{c_1(\alpha_K)=0}\quad \text{and,}\quad \boxed{d_2(\alpha_K)=0}
\end{equation}
In addition for this stationary configuration using \eqref{divofsigmaandomega} we have
\begin{equation}
	(u\cdot K\cdot\mathcal{P})^\alpha=-(\hnabla^2 u\cdot\mathcal{P})^\alpha=-\frac{1}{2}\hnabla^\mu \omega_{\mu}^\alpha
\end{equation}
Using the above relations we can see that consistency of the membrane equations with the stationary configuration also requires
\begin{eqnarray}
	\boxed{d_4(\alpha_K)+d_6(\alpha_K)=d_9(\alpha_K)},\quad\boxed{d_7(\alpha_K)=d_8(\alpha_K)}
\end{eqnarray}
There are no additional constraints from stationarity on the transport coefficients of the rest of the terms in the scalar membrane equation as these manifestly vanish for the stationary configuration. 
\section{Relation with Wald Entropy formula}
In \cite{Dandekar:2019hyc} the entropy current obtained from the membrane equation for EGB gravity was shown to match the Wald entropy of the corresponding black hole when evaluated on stationary configurations. So far we have not explicitly imposed matching of stationary membrane entropy with Wald entropy as a constraint. In this section we show that for entropy obtained from the minimal entropy current for stationary membrane configurations to match with the Wald entropy of corresponding black holes will require the imposition of constraints on the transport coefficients of the membrane equations. Later we will show how we can get around this constraint by appropriately modifying the entropy current in a simple manner. 

We now illustrate how this matching condition imposes apparent precise constraints on the transport coefficient of membrane equation in the context of EGB gravity perturbative in GB parameter. Before proceeding we write down the map between the $\alpha_K$ variable of this paper and the scaled EGB parameter of \cite{Saha:2018elg, Dandekar:2019hyc}
$$\alpha_K^2=\frac{\beta\mathcal{K}^2}{D^2}.$$

The schematic form of the sub-leading order in 1/D scalar membrane equation for EGB gravity up to $\mathcal{O}(\beta^2)$ can be written as
\begin{equation}
\nabla\cdot u=a(\alpha_K)+\frac{2}{\mathcal{K}}\left(1-\frac{\beta \mathcal{K}^2}{D^2}+q \frac{\beta^2\mathcal{K}^4}{D^4}\right)\sigma_{\alpha\beta}\sigma^{\alpha\beta}+4 \frac{\beta\mathcal{K}^2}{D^2}\left(1+ p\frac{\beta\mathcal{K}^2}{D^2}\right)\frac{\sigma_{\alpha\beta}K^{\alpha\beta}}{\mathcal{K}}
\end{equation}
where, $p$ and $q$ are unknown coefficients to be determined by explicit evaluation of the membrane equation upto $\mathcal{O}(D^{-1})$. Since, we are interested in the leading order in $1/D$ membrane entropy, it can be checked that the explicit form of the rest of the membrane equation is not important as they contribute to entropy at higher order in $1/D$.

Using the convention from the previous sections the relevant quantities with this membrane equation are
\begin{eqnarray}
b(\alpha_K)&=& \frac{4\beta\mathcal{K}^2}{D^2}\left(1+p\frac{\beta\mathcal{K}^2}{D^2}\right)\nonumber\\
W(\alpha_K)&=&\int \frac{1}{4}\frac{b}{\mathcal{K}}dK=\frac{\beta \mathcal{K}^2}{2D^2}+p\frac{\beta^2\mathcal{K}^4}{4D^4}\nonumber\\
\tilde{W}(\alpha_K)&=&\frac{1}{4}\int\frac{b}{\mathcal{K}}\left(\int\frac{b}{\mathcal{K}}dK\right)dK=\frac{\beta^2\mathcal{K}^4}{2D^4}+\mathcal{O}(k^3)
\end{eqnarray}
From the algorithm mentioned earlier the entropy current relevant up to $\mathcal{O}(\beta^2)$ is given by
\begin{eqnarray}
J^\mu_S&=&\frac{u^\mu}{4}\left(1+4 W+4\tilde{W}\right)-\frac{b(\alpha_k)}{4}\frac{(u\cdot K)^\mu}{\mathcal{K}}(1+4W)+\mathcal{J}^\mu\nonumber\\
&=&\frac{u^\mu}{4}\left(1+2\frac{\beta \mathcal{K}^2}{D^2}+ (p+2)\frac{\beta^2\mathcal{K}^4}{D^4}\right)-\frac{\beta\mathcal{K}}{D^2}\left(1+(p+2)\frac{\beta\mathcal{K}^2}{D^2}\right)(u\cdot K)^\mu+\mathcal{J}^\mu\nonumber\\
\end{eqnarray}
The piece `$\mathcal{J}^\mu$' contains parts of the current which are proportional to $j^\mu_a$, $P^\mu$ and the derivative of the function $W$. These pieces are not very relevant for the leading order expression of the entropy. The divergence of the above current accurate up to $\mathcal{O}(\beta^2)$ is given by
\begin{equation}
\hnabla\cdot J_S=\left(1+\frac{\beta\mathcal{K}^2}{D^2}+(p+q)\frac{\beta^2\mathcal{K}^4}{D^4}\right)\frac{\sigma_{\alpha\beta}\sigma^{\alpha\beta}}{\mathcal{K}}+\mathcal{O}(\beta^3)
\end{equation}
Both the expression of the entropy current and the divergence of the entropy current are consistent with the corresponding expressions obtained up to order $\mathcal{O}(\beta)$ in \cite{Dandekar:2019hyc}. We will now compute the entropy obtained by integrating the entropy current above over space-like slices of a stationary membrane configuration upto $\mathcal{O}(\beta^2)$. 
We choose a coordinate system in which the induced metric on the membrane world-volume of a stationary configuration is given by (see section $6$ of \cite{Dandekar:2019hyc})
\begin{equation}\label{killing_metric}
ds^2=h_{\mu\nu}dx^\mu dx^\nu=-e^{2\sigma(x)}(dt+a_i(x)dx^i)^2+f_{ij}(x)dx^i dx^j
\end{equation}
The velocity field of the membrane is proportional to the killing vector of the above space-time and is given by
$$u=e^{-\sigma(x)}\partial_t.$$ And the entropy obtained by integrating the entropy current along $t=\text{constant}$ slices ($\Sigma_t$)is given by (the result is independent of the particular slice since in stationary configuration the entropy current is conserved)
\begin{equation}
S_{mem}=\int_{\Sigma_t}\sqrt{h}J_S^\mu q_\mu=\int_{\Sigma_t}\sqrt{f}e^{\sigma(x)}J^t_S
\end{equation}
where, $q_\mu dx^\mu$ is the unit normal to the $\Sigma_t$ slice of the membrane space-time. Given the entropy current that we wrote above for EGB gravity above, we find the membrane entropy to be given by
\begin{eqnarray}
S_{mem}&=&\int_{\Sigma_t}\sqrt{f}\Bigg(\frac{1}{4}\left(1+2\frac{\beta\mathcal{K}^2}{D^2}+(p+2)\frac{\beta^2\mathcal{K}^4}{D^4}\right)-\frac{\beta\mathcal{K}}{D^2}\left(1+(p+2)\frac{\beta\mathcal{K}^2}{D^2}\right)K_t^t\Bigg)\nonumber\\
&=&\int_{\Sigma_t}\sqrt{f}\Bigg(\frac{1}{4}+\frac{\beta}{2D^2}\left(\mathcal{K}^2-2\mathcal{K}K_t^t\right)+(p+2)\frac{\beta^2\mathcal{K}^2}{4D^4}\left(\mathcal{K}^2-4 \mathcal{K}K_t^t\right)\Bigg)\nonumber\\
&=&\int_{\Sigma_t}\sqrt{f}\Bigg(\frac{1}{4}+\frac{\beta}{2D^2}\left(\mathcal{R}-2\mathcal{R}_t^t\right)+(p+2)\frac{\beta^2\mathcal{R}}{4D^4}\left(\mathcal{R}-4 \mathcal{R}_t^t\right)\Bigg)
\end{eqnarray}
Where, in the last line we have expressed the extrinsic curvature in terms of the intrinsic curvature of the induced metric using the Gauss-Codacci equations. All the terms above are $\mathcal{O}(D^0)$ or smaller. In \cite{Dandekar:2019hyc} it was shown that
\begin{eqnarray}
\mathcal{R}&=&\tilde{\mathcal{R}}-2\nabla^2\sigma+\mathcal{O}(D^0)\nonumber\\
\mathcal{R}_t^t&=&-\nabla^2\sigma+\mathcal{O}(D^0)\nonumber\\
\end{eqnarray}
Where, $\tilde{R}$ is the intrinsic curvature of the metric $f_{ij}dx^idx^j$. Hence,
\begin{equation}\label{mem_ent}
S_{mem}=\int_{\Sigma_t}\sqrt{f}\left(\frac{1}{4}+\frac{\beta}{2D^2}\tilde{R}+(p+2)\frac{\beta^2}{4D^4}(\tilde{R}-\nabla^2\sigma)(\tilde{R}+2\nabla^2\sigma)\right)
\end{equation}
In section $6.1$ of \cite{Dandekar:2019hyc} the Wald entropy for stationary black hole to leading order in large $D$ was shown to be given by
\begin{equation}\label{wald}
S_{Wald}=\frac{1}{4}\int_{\Sigma_v}\sqrt{\mathcal{P}}\left(1+\frac{2\beta}{D^2}\mathcal{R}_\mathcal{P}+\mathcal{O}(1/D)\right)
\end{equation}
where, $\Sigma_v$ is a spatial slice of the horizon at constant $v$ where $\partial_v$ is the generator of the event horizon. In a gauge in which the metric corrections along the $O_\mu dx^\mu$ directions vanish, we have metric on the horizon of the black hole to be given by 
\begin{equation}
g_{MN}=\eta_{MN}+O_M O_N+H^{(T)}_{MN}+\frac{1}{D}H^{(Tr)}\mathcal{P}_{MN}
\end{equation}
where, $$\mathcal{P}_{MN}=\eta_{MN}+u_M u_N+n_M n_N$$
is the projector orthogonal to the $n$ and $u$ vectors. The determinant of this projector appears in the definition of the Wald entropy written above. The expression \eqref{wald} only assumes that the metric correction are at $\mathcal{O}(1/D)$ and assumes no information about the $\beta$ dependence of the metric corrections. The projector $\mathcal{P}_{\mu\nu}$ is independent of $\beta$ and hence the Ricci scalar computed with the projector as a metric is also independent of $\beta$. 

Comparing \eqref{mem_ent} and \eqref{wald} we see that both $\Sigma_t$ and $\Sigma_v$ are $D-2$ dimensional spatial slices. The $\mathcal{O}(\beta^0)$ entropy obtained from the membrane picture matches with the Hawking area formula if in a  suitable coordinate system if we have $$f_{ij}=\mathcal{P}_{ij}.\footnote{Both $f_{ij}$ and $\mathcal{P}_{ij}$ are by construction independent of $\beta$ and hence this relation need not contain a $\beta$ dependence at higher orders}$$ This is a valid assumption as both these manifolds are smooth spatial manifolds of dimension $D-2$. 

Once, we assume that the metrics on the $D-2$ dimensional spatial slices are same, it follows that the corresponding Ricci scalar curvatures are also same and hence, there is a match to leading order in large $D$ and linear order in $\beta$. But we see that there are no $\mathcal{O}(\beta^2)$ pieces in the gravity expression \eqref{wald} and hence for the match between the membrane and Wald entropy at leading order  seems we must have $p=-2.$
Hence, it seems matching with Wald entropy can put additional constraints on the transport coefficients. 
\subsection{Apparent Tension between second law and matching with Wald entropy}
The analysis in the last section can be carried out systematically when we consider $\alpha_K$ to be a very small quantity and we can do a perturbative expansion around $\alpha_K=0$. We also see that there seems to be a tension between matching of the minimal entropy current with Wald entropy for the stationary configuration and consistency with second law for dynamical configurations. 

To see this we observe that to get a leading order expression of entropy different from the the expression of the area of the horizon we need to have a non-zero value of $b(\alpha_K)$.~Also, having a non-zero value of $b(\alpha_K)$ while doing an analysis non-perturbative in $\alpha_K$ will lead to inconsistency with second law as we saw earlier.This hints towards an apparent inconsistency between second law and matching with  Wald entropy formula at stationary configuration, if we use the minimal membrane entropy current.

We see from the perturbative  analysis of EGB gravity of \cite{Dandekar:2017aiv} that $b(\alpha_K)$ is non-zero and hence we expect this to be true even non-perturbatively. Hence we expect that the appearance of a non-zero $b(\alpha_K)$ is a common phenomena for general theories of gravity and hence the apparent inconsistency will hold for a wide class of classical theories of gravity. In the next section we will show how a small modification of the minimal entropy current can ameliorate this problem. 
\subsection{A modification of the "minimal"  membrane entropy current}
Let us define a new entropy current which has a non-trivial $\mathcal{O}(D^0)$ coefficient $T(\alpha_K)$ in front of $\frac{u^\mu}{4}$ (such that $\lim_{\alpha_K\rightarrow0}T(\alpha_K)=1$), given by
\begin{equation}
	J_n^\mu=\frac{T(\alpha_K)}{4}u^\mu+\mathcal{F}^\mu_n(\alpha_K)
\end{equation}
Hence, the divergence of the entropy current is given by
\begin{eqnarray}
	\hnabla\cdot J_n&=&\frac{T(\alpha_K)}{4}\hnabla\cdot u+\frac{1}{4}u\cdot\hnabla\left( T(\alpha_K)\right)+\hnabla\cdot \mathcal{F}^\mu_n\nonumber\\
	&=& \frac{T(\alpha_K)}{4}\hnabla\cdot u+\frac{\mathcal{K}\partial_{\mathcal{K}}(T(\alpha_K))}{4}\frac{u\cdot\hnabla \mathcal{K}}{\mathcal{K}}+\hnabla\cdot \mathcal{F}_n^\mu
\end{eqnarray}
So, from the form of the above equation it is easy to see that the new counter-term $\mathcal{F}^\mu_n$ which gives rise to a positive definite divergence of the new membrane entropy current, is related to the old counter-term $\mathcal{F}^\mu$ by the following redefinitions of the coefficients of various terms in $\mathcal{F}^\mu$
\begin{eqnarray}
	&&A_2(\alpha_K)\rightarrow \bar{A}_2(\alpha_K)=A_2(\alpha_K)T(\alpha_K)+\mathcal{K}\partial_{\mathcal{K}}(T(\alpha_K))\nonumber\\
	&& c_i(\alpha_K)\rightarrow \bar{c}_i(\alpha_K)=c_i(\alpha_K) T(\alpha_K)\quad \text{for}\quad i={1,2}\nonumber\\
	&& b_i(\alpha_K)\rightarrow\bar{b}_i(\alpha_K)= T(\alpha_K) b_i(\alpha_K)\quad \text{for,}\quad i={1,2,3}\nonumber\\
	&& d_i(\alpha_K)\rightarrow\bar{d}_i(\alpha_K) =T(\alpha_K)d_i(\alpha_K)\quad \text{for,} i={1,\ldots, 12}\nonumber\\
	\text{and,}&& b(\alpha_K)\rightarrow\bar{b}(\alpha_K)=T(\alpha_K) b(\alpha_K)
\end{eqnarray}
where, $A_2, b_i, c_i, d_i$ and $b(\alpha_K)$ are the transport coefficients of the scalar membrane equation. 

The equality type constraints on the transport coefficients coming from demanding the existence of stationary membrane configurations remain the same. The constraint from positive definite entropy production gets modified to
$$\boxed{T(\alpha_K)d_{10}(\alpha_K)\ge 0}.$$

It is easy to see that for perturbative analysis of EGB gravity we can choose $T(\alpha_K)=\left(1-(s+2)\frac{\beta^2\mathcal{K}^2}{D^4}\right)$ to make sure that the stationary membrane entropy evaluated with the membrane entropy current matches with the Wald entropy of the corresponding black hole upto $\mathcal{O}(\beta^2)$. The value of $T(\alpha_K)$ can be modified order by order in $\beta$ to take into account the transport coefficient of the scalar membrane equations at higher order in $\beta$, so that the matching with Wald entropy continues to hold.  
\subsubsection*{Matching with Wald entropy non-perturbatively in $\alpha_K$}
Let us assume that we have a theory of gravity at hand that produces a membrane equation with $b(\alpha_K)\ne 0$. In addition let us assume that for this theory of gravity satisfies all other consistency constraints on the membrane transport coefficients. This makes sure that the membrane entropy current satisfies the local form of the second law. 

We then observe that the coefficient of $u^\mu$ in the membrane entropy current determines the entropy evaluated for stationary configurations. In absence of transport coefficient $b(\alpha_K)$ this coefficient is completely determined by $T(\alpha_K)$. Hence, we can fix this coefficient by demanding that the stationary membrane entropy matches the Wald entropy of the dual black hole. 

\section{Conclusions and Outlook}
In this paper we wrote down the most general membrane equations up to sub-leading order in $1/D$ for a general class of classical gravity theories. The gravity theories we considered have a smooth limit to Einstein-Hilbert theory and also the solutions that we consider have a smooth limit to the solutions of Einstein-Hilbert theory. For these membrane equations we write down a membrane entropy current that has a positive definite divergence and hence satisfies a local form of second law of thermodynamics provided we impose certain constraints on the transport coefficients in the membrane equations. This entropy current has the least number of terms while imposing the least number of constraints on the transport coefficient of terms in the membrane equations.

We also imposed additional constraints on the transport coefficients from the existence of stationary membrane solutions dual to stationary black holes. We find that the minimal entropy current derived by us cannot at the same time satisfy the second law and also make sure that the stationary membrane entropy matches with corresponding black hole Wald entropy. We then demonstrated that a simple modification of the minimal entropy current can make sure that the stationary membrane entropy matches the Wald entropy, while making sure that the second law is not violated. 

There are a set of generalisations and checks that we wish to pursue as future projects. The most pressing thing to do will be to work out the scalar membrane equation for EGB gravity non-perturbatively in $\alpha_K$ upto sub-leading order to see how the proposal of this paper pans out. The most interesting part will be to find the coefficient of the $\sigma.K^{(TT)}$ term in the scalar membrane equation and check for regimes in higher derivative coupling where one can still have a consistent second law

It will also be of interest to us to generalise this analysis to membrane equations with charge or possibly more general matter to see what sort of constraints need to be imposed on the matter sector beyond null energy conditions for consistency with second law. Also, we will be able to see how the constraints on the membrane transport coefficients get modified in presence of matter.  
\section*{Acknowledgement}
The author would like to thank Yogesh Dandekar and Shiraz Minwalla for many insightful discussion on second law for black holes and large $D$ membranes in general. The work of the author is supported by the Ambizione grant no.  $PZ00P2\_174225/1$   of the Swiss National Science Foundation (SNSF) and partially by the NCCR grant no. $51NF40-141869$  “The Mathematics of Physics” (SwissMap)
\section*{Appendices}
\appendix
\section{$F(\psi)$ has to satisfy ODE for leading ansatz to solve gravity equation}\label{Fequation}
The expressions for the various internal curvatures of the ansatz metric at leading order in large $D$ are given by
\begin{eqnarray}\label{lead_riemann}
	\tilde{R}^A_{BCD}&=&\frac{N^2 F''}{2}\bigg(n_D n_{(B} O^A O_{C)}-n^A O_Bn_{[D} O_{C]}- n_C O^A n_{(B}O_{D)}\bigg)\nonumber\\
	&&+\frac{N^2 ((F')^2+ F F'')}{2} O^A O_B n_{[C}O_{D]}+\mathcal{O}(D)\nonumber\\
	\tilde{R}_{BD}&=&-\frac{N^2 F''+N F'\mathcal{Q}}{2}\bigg(n_{(B} O_{D)}-O_B O_D\bigg)+\frac{N^2(F')^2+N F F'\mathcal{Q}}{2}O_B O_D+\mathcal{O}(D),\nonumber\\
	\tilde{R}&=&-N^2 F''-2 N F'\mathcal{Q}-F \mathcal{Q}^2+\mathcal{O}(D)
\end{eqnarray}
where, $$N=\sqrt{d\psi\cdot d\psi},\quad F'=\partial_\psi F(\psi)\sim\mathcal{O}(D)\quad\text{and,}\quad F''=\partial^2_\psi F(\psi)\sim\mathcal{O}(D^2)$$
$$\text{and,}\quad  \mathcal{Q}=\nabla^M (\frac{\partial_M\psi}{N})\sim \mathcal{O}(D)$$
Also, we have $$\nabla^{2}\bigg(\psi^{-(D-3)}\bigg)=\mathcal{O}(D)\implies \nabla^2\psi=\psi d\psi.d\psi ~D+\mathcal{O}(D^0),$$  using which we can write
$$\mathcal{Q}=\psi N D+\mathcal{O}(D^0)$$

From the above expressions we see that all internal curvature tensors evaluate to $\mathcal{O}(D^2)$ quantities at the leading order in large $D$. For two derivative vacuum Einstein equations  $\tilde{R}_{MN}=0$ and $\tilde{R}=0$ require that the function $F(\psi)$ satisfy the following differential equation
\begin{eqnarray}
	-F''-2  F'\psi  D-F \psi^2 D^2=\mathcal{O}(D)
\end{eqnarray}
Since, in the large $D$ limit $F$ has non-trivial profile only in the region where $\psi-\sim \mathcal{O}(1/D)$ we choose to work with a new coordinate $R$ given by $\psi=1+\frac{R}{D}$ and the above differential equation to leading order in $1/D$ reduces to
\begin{equation}
	-F''(R)-2 F'(R)-F(R)=\mathcal{O}(1/D)
\end{equation}
The solution to the above equation which satisfies the boundary conditions mentioned earlier is $F(R)=e^{-R})$ which when written in terms of $\psi$ becomes $F(\psi)=\frac{1}{\psi^D}$ which matches with the expression of $F$ in the ansatz metric for Einstein-Hilbert gravity (see e.g \cite{Dandekar:2016fvw}). Hence, for Einstein gravity we have proved our assertion that if $F$ satisfies the same equation as static black holes in coordinates $R$ then ansatz metric continues to solve gravity equation at leading order. 

The gravity equations are diffeomorphism invariant and hence all the terms in ${E}^{(2m+2)}_{MN}$ have to be constructed out the internal curvatures, covariant derivatives acting on internal curvatures or all possible products of these objects with a maximum of $2m+2$ derivatives acting on the metric. In addition the factor of $D^{-2m}$ factor in front of $E_{MN}^{(2m+2)}$ in the gravity equation requires that for the ansatz metric $E_{MN}^{(2m+2)}$ evaluates to an $\mathcal{O}(D^{2m+2})$ quantity in order to contribute to the gravity equation at leading order. 

Let us assume that $E_{MN}^{(2m+2)}$ consist of $a$ internal curvature tensors ( or scalars) and hence $2m+2-2a$ derivatives acting on the internal curvatures. Since all curvatures are $\mathcal{O}(D^2)$ to leading order, the action of $2(m+1-a)$ derivatives on these internal curvatures must produce as extra factor of $D^{2m+2-2a}$. There are two ways for derivatives to act so that each derivatives produces a factor of $D$ 
\begin{enumerate}
	\item Act on $F$ or its derivatives. In this case each derivative increases order by one factor of $D$. All of these derivatives produce a factor of $N$ since $$\partial_M F=N n_M F'=\frac{\mathcal{Q}}{D\psi}$$
	\item Due to the large isometry in the dynamics, the covariant derivatives act by contraction on $n_M$ factors in the ansatz metric or its derivatives and produce factors of $\mathcal{Q}$ as
	$$\nabla_M n^M=\mathcal{Q}$$
\end{enumerate}
Hence, the gravity equation in the region $\psi-1\sim \mathcal{O}(1/D)$ become ordinary non-linear differential equation on $F$ with $\mathcal{Q}$ dependent factors. For static black hole $\mathcal{Q}=\frac{D}{r_h}+\mathcal{O}(D^0)$.  The differential equation in the static black hole solution also will be obtained in a similar manner with the extra factors of $D$ from the second type above coming from the christoffel symbols for systems with large spherical isometry. Hence, the function $F$ has to satisfy the same equation as static black holes but in $R$ coordinates. 
\section{Details of the classification of objects at various orders in $1/D$ in gravity equation}\label{classificationdetails}
\subsection{Objects in first sub-leading order in $1/D$ gravity equations}
At this order the gravity equation to evaluate to $\mathcal{O}(D)$ and hence we have
\begin{eqnarray}
	x+y+w+2z=1
\end{eqnarray}
The above equation implies that $$z=0\quad \text{and,}\quad x+y+w=1.$$
Hence, the lowest order in $D$ that any of the $\tilde{R}$s can evaluate to is $\mathcal{O}(D)$. The objects at at the different allowed orders in $D$ of $\tilde{R}$ are acting on the leading order ansatz are
\begin{itemize}
	\item at $\mathcal{O}(D^2)$:  $F$ and its first and second derivatives and $\mathcal{Q}=\nabla.n$ (see appendix \eqref{Fequation})
	\item at $\mathcal{O}(D)$: From the structure of ansatz metric, at this order $\tilde{R}$ can have $F$ and its first derivative, $\mathcal{Q}$ or $\nabla^2A_M^i$. 
\end{itemize}
Hence, the various objects contributing to the gravity equation at this order are
\begin{enumerate}
	\item $x=1,y=w=0$. All curvatures evaluate to $\mathcal{O}(D^)$ and all derivatives contribute $D$ each. Action of derivatives on $F$ does not produce objects as we defined them. The action of derivatives by contraction on leading order $\tilde{R}s$ produce factors of $\mathcal{Q}$ and hence, action of the $x=1$ derivative on leading order curvature produce the objects 
	$$\nabla_M \mathcal{Q} \quad \text{and,}\quad \nabla_M A_N^{(i)}.$$
	From here on we will suppress the indices unless it interferes with clarity. 
	\item $w=1, x=0=y$. Following similar logic the objects from here are of type
	$$\nabla A$$
	\item $y=1, x=w=0$ gives $$\nabla^2 A, \nabla^2 Q$$
\end{enumerate}
hence, the list of all objects that can contribute to gravity equation at first sub-leading order are
\begin{equation}
	\boxed{	1)\nabla_M\mathcal{Q}\quad 2) \nabla_M A_N^{(i)}\quad 3)\nabla^2 A_M^{(i)}\quad 4)\nabla^2\mathcal{Q}}
\end{equation}
Taking into account the above objects the first order corrected metric will have the schematic form
\begin{equation}
	\eta_{MN}+F(\psi)O_M O_N+\frac{1}{D}\bigg(F_1(\psi)A_N^{(i)}\nabla_M Q+F_2(\psi) \nabla_M A_N^{(i)}+F_3(\psi)\nabla^2 A_M^{(i)}+F_4(\psi) A_M^{(i)}A_N^{(j)}\nabla^2\mathcal{Q}\bigg)
\end{equation} 
\subsection{Objects at second sub-leading order in 1/D}
The gravity equation at this order should evaluate to $\mathcal{O}(D^0)$ and hence we should have
$$x+y+w+2z=2.$$ 
One of the possible solutions is $z=1, x=y=w=0$. So, the minimum order in $D$ the curvatures can evaluate to is $\mathcal{O}(D^0)$. Also, the maximum number of external derivatives that we can have is given by the solution $y=2, x=w=z=0$. This corresponds to two factors of $\nabla^2$.

We write down the objects at various orders in $D$ in $\tilde{R}$s here. While writing down we have used the fact that the $\nabla$s are covariant derivatives w.r.t the flat space-time and hence they commute. We write down the contributions ( suppressing the indices) by inspection as  
\begin{itemize}
	\item at $\mathcal{O}(D^2)$ the contributions only come from ansatz part and they are same as above.
	\item at $\mathcal{O}(D)$ the objects are
	$$\nabla A, \nabla^2 A, \nabla Q, \nabla^2 Q$$
	\item at $\mathcal{O}(D^0)$ the objects are
	$$\nabla\nabla A, \nabla\nabla \mathcal{Q}, \nabla\nabla^2 A, \nabla\nabla^2 \mathcal{Q}, \nabla \mathcal{Q}, \nabla^2 \mathcal{Q}, \nabla^2\nabla^2 A, \nabla^2 \nabla^2\mathcal{Q}$$
	$$\nabla A \nabla\mathcal{Q}, \nabla A\nabla^2 \mathcal{Q}, \nabla^2 A\nabla\mathcal{Q},\nabla^2 A\nabla A, \nabla^2 A\nabla^2A, \nabla^2A\nabla^2\mathcal{Q},\nabla A\nabla A$$
	Objects in the first line are linear in membrane vectors $A_m^{(i)}$ and $\mathcal{Q}$ and objects in second line are bi-linear.
\end{itemize}
In the above whenever $\nabla$ appears (not as $\nabla^2$), it means it comes with a free index. The same goes for $A_M^{(i)}$. 

Next, let us consider the various solution of $x+y+w+2z=2$ case by case
\begin{enumerate}
	\item $z=1, x=y=w=0$: One $\tilde{R}$ at $\mathcal{O}(D^0)$ and rest of them at $\mathcal{O}(D^2)$. And all derivatives act either on $F$s or as divergences on tensors. The list of object is all the objects at $\mathcal{O}(D^0)$ in $\tilde{R}$ above. 
	\item $x+y=2, w=z=0$. This has three possibilities
	\begin{itemize}
		\item $x=2,y=0=w=z$, the objects from this type are
		$$\nabla\nabla\mathcal{Q}, \nabla\nabla A, \nabla A\nabla A, \nabla \mathcal{Q}\nabla\mathcal{Q}, \nabla A\nabla\mathcal{Q}$$
		\item $y=2,x=w=z=0$
		$$\nabla^2\nabla^2\mathcal{Q}, \nabla^2\nabla^2 A, \nabla^2\mathcal{Q}\nabla^2 A, \nabla^2\mathcal{Q}\nabla^2\mathcal{Q}$$
		\item $x=y=1, z=w=0$
		$$\nabla\mathcal{Q}\nabla^2\mathcal{Q},\nabla\mathcal{Q}\nabla^2 A, \nabla^2\mathcal{Q}\nabla A, \nabla A\nabla^2 A$$ 
	\end{itemize}
	\item $x+w=2, y=z=0$. There are two new possibilities
	\begin{itemize}
		\item $x=1, w=1$
		$$\nabla\mathcal{Q},\nabla A,\nabla^2\mathcal{Q},\nabla^2 A$$
		\item $w=2, x=0$, All bi-linears of
		$$\nabla\mathcal{Q},\nabla A,\nabla^2\mathcal{Q},\nabla^2 A$$
	\end{itemize}
	\item $y+w=2$ with the only new possibility of $y=1, w=1$
	$$\nabla^2\nabla\mathcal{Q}, \nabla^2\nabla A, \nabla^2\nabla^2\mathcal{Q},\nabla^2\nabla^2 A$$
\end{enumerate}
Hence, the exhaustive list of objects that can contribute at this order is given by
\begin{eqnarray}
	&&\boxed{\nabla\nabla A,\quad \nabla\nabla \mathcal{Q},\quad \nabla\nabla^2 A,\quad \nabla\nabla^2 \mathcal{Q},\quad \nabla \mathcal{Q},\quad \nabla^2 \mathcal{Q},\quad \nabla^2\nabla^2 A, \quad\nabla^2 \nabla^2\mathcal{Q},\quad \nabla A, \quad \nabla^2 A}\nonumber\\
	&&\boxed{\nabla A \nabla\mathcal{Q}, \quad\nabla A\nabla^2 \mathcal{Q},\quad \nabla^2 A\nabla\mathcal{Q},\quad \nabla^2 A\nabla A, \quad\nabla^2 A\nabla^2A,\quad \nabla^2A\nabla^2\mathcal{Q},\quad\nabla A\nabla A,\quad \nabla\mathcal{Q}\nabla\mathcal{Q},}\nonumber\\
	&&\boxed{ \nabla^2\mathcal{Q}\nabla^2\mathcal{Q},\quad \nabla^2\mathcal{Q},\nabla\mathcal{Q}}
\end{eqnarray}
Or in other words the maximum derivative objects for linear in $A_M^{(i)}$ and $\mathcal{Q}$ are $$\nabla^2\nabla^2 \mathcal{Q}\quad \text{and,}\quad \nabla^2\nabla^2A$$ . And the maximum derivative objects which are bi-linear are
$$\nabla^2\mathcal{Q}\nabla^2\mathcal{Q},\quad \nabla^2 A \nabla^2 A,\quad \text{and,}\quad\nabla^2\mathcal{Q}\nabla^2 A.$$  
There is also a constraint on three derivative objects wherein two of the derivatives have to act as $\nabla^2$. This is true for both the linear and bi-linear objects. All lower order derivative objects can appear in the gravity equation and are unconstrained. 
\section{Some details of the derivation of  objects in membrane equation from objects in gravity equation at first sub-leding order}\label{gravtomemeqn}
Let us first consider the objects with the maximum number of derivatives on the membrane vectors $A_M^{(i)}$ and $\mathcal{Q}$ in the gravity equations
$$\nabla^2 n_M,\quad\nabla^2 u_M\quad \text{and,}\quad \nabla^2 \mathcal{Q}$$
Using the subsidiary conditions on normal derivative of normal and velocity vector we can write
\begin{eqnarray}
	\nabla^2 u_M&=&\left(P^{BC}+n^B n^C\right)\nabla_C\left((P_B^D+n_B n^D)\nabla_D u_M\right)\nonumber\\
	&=&(P\cdot \nabla)^B (P\cdot \nabla)_B u_M+n^B n\cdot \nabla ((P\cdot \nabla)_B u_M)\quad (\because n\cdot \nabla n_M=0=n\cdot \nabla u_M)\nonumber\\
	&=&(P\cdot \nabla)^2 u_M+\mathcal{O}(D^0)
\end{eqnarray}
where, $P_{AB}=\eta_{AB}-n_A n_B$. We keep only the terms leading order in $D$ throughout as that is the relevant piece that contributes objects to membrane equation.  $(P\cdot \nabla)^2 u_M$ can be thought of as the Laplacian acting on the velocity field along the transverse direction on $\psi=1$ surface and hence is $\mathcal{O}(D)$ due to the large isometry. Similarly,
\begin{eqnarray}
	\nabla^2 n_M =&=& \left(P^{BC}+n^B n^C\right)\nabla_C\nabla_B\nabla_M B\nonumber\\
	&=& P^{BC} \nabla_M \nabla_C \nabla_B B+n^C \nabla_C(n^B \nabla_M \nabla_B B) \quad (\because n\cdot\nabla n_M=0)\nonumber\\
	&=& \nabla_M (P^{BC} \nabla_b\nabla_C B)-\nabla_M P^{BC} \nabla_C \nabla_B B\nonumber\\
	&=&\nabla_M \mathcal{Q}+\mathcal{O}(D^0)
\end{eqnarray}
Next we consider the objects with lower number of derivatives given by $$\nabla_A n_B,\quad \nabla_A u_B,\quad \text{and,}\nabla_A \mathcal{Q}$$
Using our subsidiary conditions we have on the membrane surface
$$\nabla_A n_B=\nabla_A \nabla_B B$$
which is the extrinsic curvature of any member of the family of surface we consider. There is no anti-symmetric data in $\nabla_A n_B$ with our choice of subsidiary conditions.

Rewriting $\nabla_A u_B$ using the projectors orthogonal to $n_A$ we get
\begin{equation}
	\nabla_A u_B=(P.\nabla u.P)_{AB}+P_A^M \nabla_M u_N n^N n_B=(P\cdot\nabla u\cdot P)-n_Bu^M\nabla_M\nabla_A B (\because u\cdot n=0)
\end{equation}
The last term carries the information of the the extrinsic curvature again and hence the new piece of information contained in $\nabla_A u_B$ is $P_A^C\nabla_C u_D P^D_B$.
Hence, using the subsidiary conditions we have written all the objects in the gravity equation at this order in terms of their transverse derivatives acting along the surface $\psi=1$. After this to convert this objects to world volume objects in the membrane world volume we use the map
$$P_{MN}\rightarrow g_{\mu\nu} \quad \mathcal{Q}\rightarrow \mathcal{K}\quad \text{and,}\quad \nabla\rightarrow \hnabla$$
where $\mu\nu\ldots$ are coordinates in membrane world-volume. $g_{\mu\nu}$ is the induced metric in the world volume, $\hnabla$ is the covariant derivative w.r.t $g_{\mu\nu}$ and $\mathcal{K}$ is the trace of the extrinsic curvature of the membrane.
Using this map and the above analysis the precursor to the objects contributing to the membrane equation coming from objects in first sub-leading order membrane equations are
$$\nabla^2 u_M P^M_A\rightarrow\hnabla^2u_\alpha\quad \{\nabla^2 n_A,\nabla_A\mathcal{Q}\}\rightarrow \hnabla_\alpha\mathcal{K}\quad\nabla^2\mathcal{Q}\rightarrow\hnabla^2\mathcal{K}\quad \nabla_A n_B\rightarrow K_{\alpha\beta}\quad  \nabla_M u_N\rightarrow\nabla_\alpha u_\beta$$

Defining a projector orthogonal to the membrane velocity vector in the membrane world volume by 
$$\mathcal{P}_{\alpha\beta}=g_{\alpha\beta}+u_\alpha u_\beta. $$
we can write the following tensor decomposition of $\hnabla_\mu u_\nu$ w.r.t $\mathcal{P}_{\mu\nu}$ as
\begin{eqnarray}\label{gradu}
	\nabla_\mu u_\nu&=& \sigma_{\mu\nu}+\omega_{\mu\nu}+a_\nu u_\mu+\frac{\mathcal{P}_{\mu\nu}}{D-2}\hnabla\cdot u\nonumber\\
	\text{where,}\quad \sigma_{\mu\nu}&=&\mathcal{P}^\alpha_\mu \frac{\hnabla_\alpha u_\beta+\hnabla _\beta u_\alpha}{2} \mathcal{P}^\beta_\nu-\frac{\mathcal{P}_{\mu\nu}}{D-2}\hnabla\cdot u\nonumber\\
	\omega_{\mu\nu}&=&\mathcal{P}^\alpha_\mu \frac{\hnabla_\alpha u_\beta-\hnabla _\beta u_\alpha}{2} \mathcal{P}^\beta_\nu\nonumber\\
	a_\nu&=& u\cdot \hnabla u_\nu
\end{eqnarray}
Similarly, the tensor decomposition of the extrinsic curvature $K_{\mu\nu}$ is given by
\begin{eqnarray}
	K_{\mu\nu}&=& K^{(TT)}_{\mu\nu}+u_\mu \left(u\cdot k\cdot \mathcal{P}\right)_\nu+u_\nu \left(u\cdot k\cdot \mathcal{P}\right)_\mu+u_\mu u_\nu \left(u\cdot K\cdot u\right)+\frac{\mathcal{P}_{\mu\nu}}{D-2}\mathcal{K}\nonumber\\
	\text{where,}\quad K^{(TT)}_{\mu\nu}&=&\mathcal{P}^\alpha_\mu K_{\alpha\beta}\mathcal{P}^\beta_\nu-\frac{\mathcal{P}_{\mu\nu}}{D-2}\mathcal{K}\nonumber\\
	\mathcal{K}&=&\mathcal{P}^{\mu\nu}K_{\mu\nu}
\end{eqnarray}
The tensor decomposition of $\hnabla_\alpha \mathcal{K}$ is given by
\begin{equation}
	\hnabla_\alpha \mathcal{K}=\hnabla_\mu \mathcal{K}\mathcal{P}^\mu_\alpha +u_\alpha u\cdot \hnabla \mathcal{K}
\end{equation}
The decomposition of $\hnabla^2 u_\mu$ is given by
\begin{eqnarray}
	\hnabla^2 u_\mu&=&\mathcal{P}^\alpha_\mu \hnabla^2 u_\alpha +\hnabla^2 u_\alpha u^\alpha u_\mu \nonumber\\
	&=&\mathcal{P}^\alpha_\mu \hnabla^2 u_\alpha +\hnabla_\beta\left(u^\alpha \hnabla_\beta u_\alpha\right) u_\mu-\hnabla_\beta u^\alpha \hnabla_\beta u_\alpha \nonumber\\
	&=&\mathcal{P}^\alpha_\mu \hnabla^2 u_\alpha -\hnabla_\beta u^\alpha \hnabla_\beta u_\alpha \quad (\because u\cdot u=-1)\nonumber\\
	&=& \mathcal{P}^\alpha_\mu \hnabla^2 u_\alpha +\mathcal{O}(D^0)
\end{eqnarray}
Where in the last line, we have used the property that $\nabla\cdot u=\mathcal{O}(D^0)$ in the decomposition \eqref{gradu} to observe that $\hnabla_\beta u^\alpha \hnabla u_\alpha\sim \mathcal{O}(D^0)$. 
Also, from the definition of the shear tensor $\sigma_{\mu\nu}$ and vorticity $\omega_{\mu\nu}$ introduced earlier, it is easy to see that
\begin{eqnarray}\label{divofsigmaandomega}
	&&\hnabla^\mu\sigma_{\mu\alpha}=\frac{1}{2}\Bigg(\hnabla^2u_\beta +\mathcal{K}(u\cdot K)_\beta \Bigg)\mathcal{P}^\beta_\alpha+\mathcal{O}(D^0)\nonumber\\
	&&\hnabla^\mu\omega_{\mu\alpha}=\frac{1}{2}\Bigg(\hnabla^2u_\beta -\mathcal{K}(u\cdot K)_\beta \Bigg)\mathcal{P}^\beta_\alpha+\mathcal{O}(D^0)
\end{eqnarray}

Hence, there is an intrinsic relation between $(\hnabla^\mu \sigma_{\mu\nu},\hnabla^\mu \omega_{\mu\nu})$ and $(\hnabla^2 u_\nu \mathcal{P}^\nu_\mu,\mathcal{K}u^\mu K_{\mu\nu})$ at leading order in large $D$ and wither pair can be used in the set of independent objects contributing to the leading order membrane equations.  

Hence, the list of independent vector objects that can contribute to the vector membrane equation at leading order in large $D$ are

\begin{equation}
	\boxed{1) \hnabla^2 u_\nu \mathcal{P}^\nu_\mu\quad2)\hnabla_\nu\mathcal{K}\mathcal{P}^\nu_\mu\quad3)a_\mu=u\cdot\hnabla u_\mu\quad\text{and,}\quad4)(u\cdot K\cdot \mathcal{P})_\mu}
\end{equation}
And the list of scalar objects that can contribute to the scalar membrane equation are
\begin{eqnarray}
	\boxed{1) u\cdot\hnabla\mathcal{K}\quad 2)u\cdot K\cdot u\quad 3)\hnabla^2\mathcal{K}}
\end{eqnarray}
\bibliographystyle{JHEP}
\bibliography{ssbib}
\end{document}